%% file: main.tex
\documentclass[acmsmall,screen]{acmart}

\usepackage{graphicx}
\usepackage{subcaption}
\usepackage{enumitem}
\setlist[itemize]{leftmargin=*}
\setlist[enumerate]{leftmargin=*}
\usepackage{xcolor}
\usepackage{tcolorbox}
\usepackage{multirow}
\usepackage{pifont}
\usepackage{wrapfig}
\usepackage{caption}
\usepackage[ruled,vlined]{algorithm2e}

\title{Names Are All You Need: Effective and Safe Regression Test Selection for Python}
\author{You Wang}
\affiliation{%
  \department{College of Computer Science and Technology and The State Key Laboratory of Blockchain and Data Security}
  \institution{Zhejiang University}
\city{Hangzhou}
\country{China}}
\email{prinzywang@zju.edu.cn}

\author{Michael Pradel}
\affiliation{%
  \institution{CISPA Helmholtz Center for Information Security}
  \city{Stuttgart}
  \country{Germany}}
\email{michael@binaervarianz.de}

\author{Zhongxin Liu}
\authornote{Corresponding Author}
\affiliation{%
  \department{College of Computer Science and Technology and The State Key Laboratory of Blockchain and Data Security}  
  \institution{Zhejiang University}
  \city{Hangzhou}
\country{China}}
\email{liu_zx@zju.edu.cn}

\newcommand{\appname}{{\textsc{NameRTS}}\xspace}

\begin{document}

\begin{abstract}
Regression test selection (RTS) reduces the cost of regression testing by executing only those tests affected by a code change. 
Despite extensive study of RTS in statically typed languages such as Java, achieving effective and safe RTS in Python is challenging.
Python’s dynamic typing makes precise call-graph construction difficult, which can cause call-graph-based RTS to miss affected tests, and hence, compromise safety.
Python’s eager importing mechanism, in contrast, renders file-level dependency analysis overly conservative.
This paper presents \appname, the first Python RTS approach based on fine-grained dependency analysis. 
\appname models a Python program as a bipartite graph of code element nodes (e.g., classes, functions, global variables) and name nodes (i.e., identifiers used to reference code elements), with edges capturing definitions and references.
RTS is formulated as a reachability problem on this graph: a test is selected if any modified code element is reachable from the names used in that test.
This design avoids call-graph construction, enabling a conservative analysis amenable to safety.
To control dependency cascades introduced by coarse name matching, \appname applies two pruning strategies that leverage prior test executions and context information to refine name matching.
To evaluate \appname, we construct the first Python RTS dataset with a ground truth indicating which test files are affected by each commit.
It includes 500 commits drawn from 10 real-world Python projects.
We compare \appname with the best-performing baseline, BabelRTS, an RTS technique based on coarse file-level dependencies.
On this benchmark, \appname skips 69.90\% of test files on average, outperforming BabelRTS by 146.5\%.
It also reduces end-to-end testing time by 45.59\%, yielding a 107.7\% improvement over BabelRTS.
In terms of safety, \appname selects all affected tests for 99.6\% of commits, with only rare misses in exceptional cases.
In contrast, BabelRTS is safe for 76.6\% of commits.
These results demonstrate the effectiveness of \appname, paving the way for more efficient regression testing in Python.
\end{abstract}

\maketitle

\LinesNumbered
\input{sections/introduction}

\input{sections/motivation}

\input{sections/approach}

\input{sections/evaluation}
\input{sections/discussion}

\input{sections/relatedwork}
\input{sections/conclusion}

\section{Data Availability}
Our code and data are available: \url{https://github.com/ZJU-CTAG/NameRTS}

\begin{acks}
This research/project is supported by the National Natural Science Foundation of China (No.92582107), the Fundamental Research Funds for the Central Universities (No.226-2025-00067), and the German Research Foundation (DFG; projects 492507603, 516334526, and 526259073). 
\end{acks}

\bibliographystyle{ACM-Reference-Format}
\bibliography{references}
\end{document}

%% file: sections/introduction.tex
\section{Introduction}

Regression testing is a standard practice in software maintenance. 
It checks whether code changes break existing functionality or introduce new defects, helping developers catch problems early \cite{leung1989insights,wong1997study,yoo2012regression,zhang2026can}. 
Unfortunately, regression testing is costly \cite{harrold1993methodology,chittimalli2009recomputing}. 
In continuous integration, the full test suite is often executed on every commit. 
Bouzenia et al. report that building and testing account for 91.2\% of virtual machine time on GitHub Actions~\cite{bouzenia2024resource}, highlighting the scale of this cost.
On the other hand, Győri et al. show that across 13961 change sets in an open-source ecosystem, only about 7.8\% to 17.4\% of tests are actually relevant to a typical code change \cite{gyori2018evaluating},
demonstrating that running full test suites is often an order of magnitude more expensive than necessary.

To reduce the cost of regression testing, \emph{regression test selection} (RTS) has been proposed \cite{rothermel1997safe,law2003whole,orso2004scaling}. 
Instead of executing all tests, RTS analyzes a code change and selects only those tests that are likely to be affected by the change \cite{leung1990study,engstrom2010systematic,yoo2012regression}. 
RTS techniques aim for two objectives: \emph{effectiveness}, by reducing the number of selected tests and end-to-end testing time; 
and \emph{safety}, by ensuring that all tests that could reveal a regression remain in the selection \cite{rothermel2002analyzing}. 
Most successful RTS techniques are based on dependency analysis, i.e., selecting a test if it depends on a modified component \cite{yoo2012regression}. 
To date, the majority of RTS techniques have been developed and evaluated for statically typed languages, most notably Java \cite{gligoric2015practical,liu2023more,zhang2018hybrid,zhang2024hybrid}.

Python’s flexibility and versatility have made it widely adopted across domains such as scientific computing and data analysis, and it is now considered the lingua franca of artificial intelligence~\cite{stackoverflowpython,jetbrainpython,tiobepython,jiang2025agentic,deng2025nocode}.
According to a GitHub report~\cite{githubpython}, Python was the second most popular language in 2025.
While RTS has been widely explored in statically typed languages \cite{gligoric2015practical,zhang2024hybrid}, 
achieving effective and safe RTS in Python faces two fundamental challenges.
First, Python’s dynamic typing makes it difficult to accurately infer variable types and statically resolve method definitions. 
As a result, precise and efficient call-graph construction for Python is significantly harder. 
Existing call graph construction tools for Python face notable completeness and scalability limitations. 
For example, Bouzenia et al.~\cite{bouzenia2024dypybench} compare the call graphs produced by state-of-the-art static and dynamic analyzers (PyCG~\cite{salis2021pycg} vs. DynaPyt~\cite{eghbali2022dynapyt}) and find PyCG fails on 11 of 50 real-world projects due to timeouts, memory exhaustion, or crashes, and that even on the 39 successful projects it detects only 49\% of the call edges observed by DynaPyt at runtime.
Second, Python’s eager importing mechanism executes global-scope code across modules even when only a submodule is imported.
Specifically, when a submodule is imported (e.g., \texttt{sympy.core.numbers}), the Python interpreter implicitly imports and executes the \texttt{\_\_init\_\_.py} files of all parent packages (e.g., \texttt{sympy/\_\_init\_\_.py} and \texttt{sympy/core/\_\_init\_\_.py}), which can in turn trigger additional imports and code execution \cite{pythonimport}.

These two challenges substantially limit the applicability of existing RTS techniques to Python.
Most prior RTS approaches are based on dependency analysis, which can be broadly categorized into function-level and file-level techniques.
\emph{Function-level RTS}~\cite{soetens2016change,li2019method,blondeau2017test,zhang2018hybrid,zhang2024hybrid} relies on call graph construction to approximate test reachability.
In Python, the completeness and scalability limitations of call graph construction make such approaches unsafe and ineffective.
\emph{File-level RTS}~\cite{gligoric2015practical,legunsen2016extensive,legunsen2017starts,maurina2025babelrts} constructs dependencies between source files and selects tests whose dependent files are modified.
In Python, this approach is fundamentally limited, since Python’s eager importing feature inflates dependency scopes and leaves little room for meaningful reduction.
Beyond dependency-based techniques, \emph{coverage-based RTS}~\cite{kauhanen2021regression} selects tests based on whether code changes were covered in previous executions.
Because eager importing executes substantial global code during test initialization, many tests exhibit broad coverage even without semantic dependence.
As a result, changes to the global code force the re-execution of most tests, severely limiting effectiveness.

To enable effective RTS in Python, our core insight is that while constructing a precise and complete call graph for Python is difficult, constructing an over-approximate dependency structure is comparatively easy. 
By assuming that any reference to a name may reach all code elements defined under that name, we can build a complete, though imprecise, dependency graph without requiring type inference or static call-graph resolution.

Motivated by this insight, this paper introduces \appname, the first Python RTS approach that tracks fine-grained dependencies.
\appname is built on an algorithm we call \emph{name-based dependency propagation}.
At its core, 
\appname models a Python project as a bipartite graph of name nodes and code element nodes. 
When a code element is implemented, it may reference other names that are not local variables, i.e., the names point to code elements defined outside the current one. 
We treat these references as \emph{external names}. 
In the graph, definition edges connect each name node to the code elements defined under that name, and usage edges connect each code element to the external names it references.
With this graph, \appname formulates test selection as a reachability problem.
A test is selected if and only if a reachability analysis starting from its external names reaches any modified code element.
This avoids relying on call-graph construction or type inference and handles Python’s eager importing feature effectively by tracking dependencies at the function level rather than at the file level, giving higher precision while keeping the analysis lightweight.

A downside of a purely name-based approach is that it may trigger dependency cascades: 
a name may match many code elements that are not actually reachable by the test, and these elements introduce additional names that, in turn, pull in yet more elements.
This recursive process can lead to an exponential expansion of reachable code elements, far beyond what the test could ever reach. 
To reduce this effect, \appname applies two pruning strategies. 
The first identifies names that tend to cause such expansions and prunes their code element nodes unless they were exercised by the test in previous executions.
The second applies lightweight, context-aware name-element matching to rule out code elements that are invalid targets of a given name, using the defining file and class of the referencing code element to narrow the set of reachable code elements.
Together, these pruning steps cut away paths with no evidence of being reachable by the test, improving precision with a negligible impact on safety in practice.

To evaluate \appname, we construct a dataset of 500 commits drawn from 10 open-source Python projects on GitHub, with code sizes ranging from 12k to 426k lines of code. 
For each commit, we collect the test files that depend on the changed elements by instrumenting the modified code and recording which tests invoke it at runtime, yielding the first Python RTS dataset with a ground truth.
We compare \appname against the state-of-the-art Python RTS approach BabelRTS \cite{maurina2025babelrts}. 
Since BabelRTS is not always safe, we additionally implement EkstaP, a safe file-level RTS baseline based on Ekstazi~\cite{gligoric2015practical}, which was originally proposed for Java.
Our results show that \appname substantially outperforms both baselines.
Across all projects, \appname skips 69.90\% of test files and reduces end-to-end testing time by 45.59\%.
This corresponds to improvements of 842.3\% and 1371.5\% over EkstaP, and 146.5\% and 107.7\% over BabelRTS, respectively.
In terms of safety, \appname selects all affected tests for 99.6\% of commits, whereas BabelRTS is safe on only 76.6\% of commits.

In summary, this paper makes the following contributions:
\begin{itemize}[leftmargin=1.2em, topsep=0pt, itemsep=1pt]
\item We propose \appname, the first Python RTS approach based on fine-grained dependency analysis, 
built on a novel algorithm called \emph{name-based dependency propagation}.
\item We curate the first Python RTS dataset with ground truth, covering complex, real-world projects with up to 426k lines of code.
\item We conduct an extensive empirical evaluation on this dataset, showing that \appname skips 69.90\% of test files and reduces testing time by 45.59\%, substantially outperforming state-of-the-art baselines while remaining safe on 99.6\% of commits.
\item We release an open-source package~\cite{package} to support reproducibility and future research.
\end{itemize}

%% file: sections/motivation.tex
\section{Motivating Example}

\begin{figure*}[t]
\centering

\begin{subfigure}{0.45\textwidth}
    \centering
    \resizebox{0.85\textwidth}{!}{
    \includegraphics[width=\linewidth]{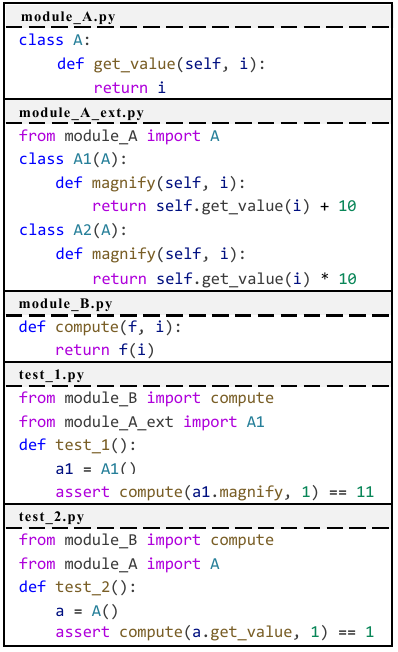}
    }
    \caption{Example project.}
    \label{fig:example_code}
\end{subfigure}
\hfill
\begin{subfigure}{0.45\textwidth}
    \centering

    \begin{subfigure}{\linewidth}
        \centering
        \includegraphics[width=0.8\linewidth]{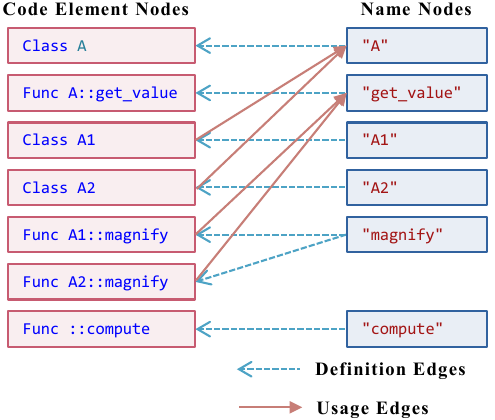}
        \caption{Bipartite name-element graph.}
        \label{fig:example_graph}
    \end{subfigure}

    \vspace{0.7em}

    \begin{subfigure}{\linewidth}
        \centering
        \includegraphics[width=\linewidth]{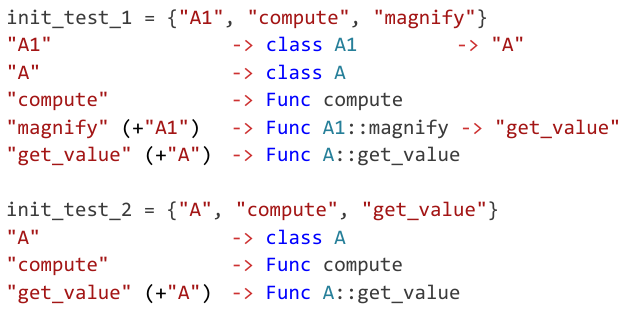}
        \caption{Propagation procedure for test files.}
        \label{fig:example_propagation}
    \end{subfigure}

\end{subfigure}

\caption{Motivating example of name-based dependency propagation.
}
\label{fig:example}
\end{figure*}

Figure \ref{fig:example} presents a project illustrating how \appname operates. 
The project contains five files. 
The test files \texttt{test\_1.py} and \texttt{test\_2.py} import the function \texttt{compute} from \texttt{module\_B.py}. 
The function \texttt{compute} takes a callable and its input value, and invokes the callable with that input, representing a function-based callback pattern.
In \texttt{test\_1.py}, the test imports class \texttt{A1} from \texttt{module\_A\_ext.py}, constructs an instance, and passes its method \texttt{magnify} together with the integer \texttt{1} to \texttt{compute}. 
Class \texttt{A1} inherits from \texttt{A}, and \texttt{A1::magnify} invokes \texttt{A::get\_value} through \texttt{self}, capturing typical object-oriented behavior and dynamic method dispatch. 
In contrast, \texttt{test\_2.py} imports \texttt{A} directly and exercises only \texttt{A::get\_value}. 
Since \texttt{magnify} has multiple implementations, i.e., in \texttt{A1} and \texttt{A2}, the example contains reachable and unreachable method definitions with the same name.

Call-graph-based RTS selects more tests than necessary in this scenario. 
For example, PyCG \cite{salis2021pycg}, a state-of-the-art static call graph builder for Python, treats the first argument of \texttt{compute} as a callable and therefore adds call edges from \texttt{compute} to multiple methods that are passed as arguments. 
In this example, the call graph contains edges from \texttt{compute} to both \texttt{A::get\_value} and \texttt{A1::magnify}. 
This means a change to \texttt{A1::magnify} selects both \texttt{test\_1.py} and \texttt{test\_2.py} even though only \texttt{test\_1.py} can reach that method at runtime.
File-level RTS faces a similar issue.
A change to \texttt{A2::magnify}, which no test executes, would force \texttt{test\_1.py} to be selected, because it imports the modified file \texttt{module\_A\_ext.py}.

Our idea is to model this example project as a bipartite graph shown in Figure \ref{fig:example_graph}.
Code element nodes represent classes, including \texttt{A}, \texttt{A1} and \texttt{A2}, as well as functions, including \texttt{A::get\_value}, \texttt{A1::magnify}, \texttt{A2::magnify}, and \texttt{compute}. 
Name nodes correspond to the names through which these code elements are referenced. 
Definition edges connect each name to the code elements defined under that name. 
Usage edges connect each code element to the external names it references. 
For example, the name \texttt{“A1”} is defined by class \texttt{A1},  
and class \texttt{A1} references \texttt{“A”} as its superclass. 
These edges capture how definitions and name references are interrelated in the program.
To determine which code elements \texttt{test\_1.py} depends on, \appname starts from external names visible in the test and performs name-based dependency propagation. 
It follows definition edges to find candidate code elements for each name, then follows usage edges to extract newly referenced names, repeating until no new names are reached. 
Figure \ref{fig:example_propagation} shows the propagation steps. 
In this example, an attribute (e.g., \texttt{A1::magnify}) is reachable only when its defining class (e.g., \texttt{A1}) is reachable.
The final reachable set consists of code elements that the test may use. 
A change to any element in this set, such as \texttt{A1::magnify}, causes \texttt{test\_1.py} to be selected for re-execution. 
A change outside this set, such as \texttt{A2::magnify}, does not.
Thus, this graph avoids the unnecessary selection that file-level RTS produces when any part of \texttt{module\_A\_ext.py} changes.
In contrast to call-graph-based RTS, \appname does not select \texttt{test\_2.py} when \texttt{A1::magnify} is modified. 

This example illustrates three key advantages of name-based dependency propagation for Python RTS.
First, the approach is lightweight and amenable to safety.
Instead of precise name resolution, \appname relies on simple name matching and conservatively considers all possible definitions of a referenced name.
Compared to call-graph construction, this over-approximate strategy avoids complex type inference, is easier to implement, and simplifies ensuring completeness and safety.
Second, despite its simplicity, the approach enables effective test selection with fine-grained dependencies.
Dependencies are tracked primarily at the function level, yielding finer granularity than file-level RTS.
Moreover, the analysis is fast enough to be performed separately for each test file, avoiding the conflation of dependencies across different tests that occurs in call-graph-based RTS, as illustrated in the motivating example above.
The same mechanism naturally extends to global variables, enabling variable-level dependency tracking and preventing unnecessary test selection caused by modifications to global variable definitions.
Third, while name-based dependency propagation is conservative and may propagate spurious dependencies, resulting in unnecessary test selection, this limitation can be significantly mitigated through targeted pruning.
\appname restricts propagation to code elements with evidence of reachability or contextual validity, which improves precision while affecting safety only in a negligible way in practice.

%% file: sections/approach.tex
\section{Approach}
\begin{figure*}
    \centering
    \includegraphics[width=0.98\linewidth]{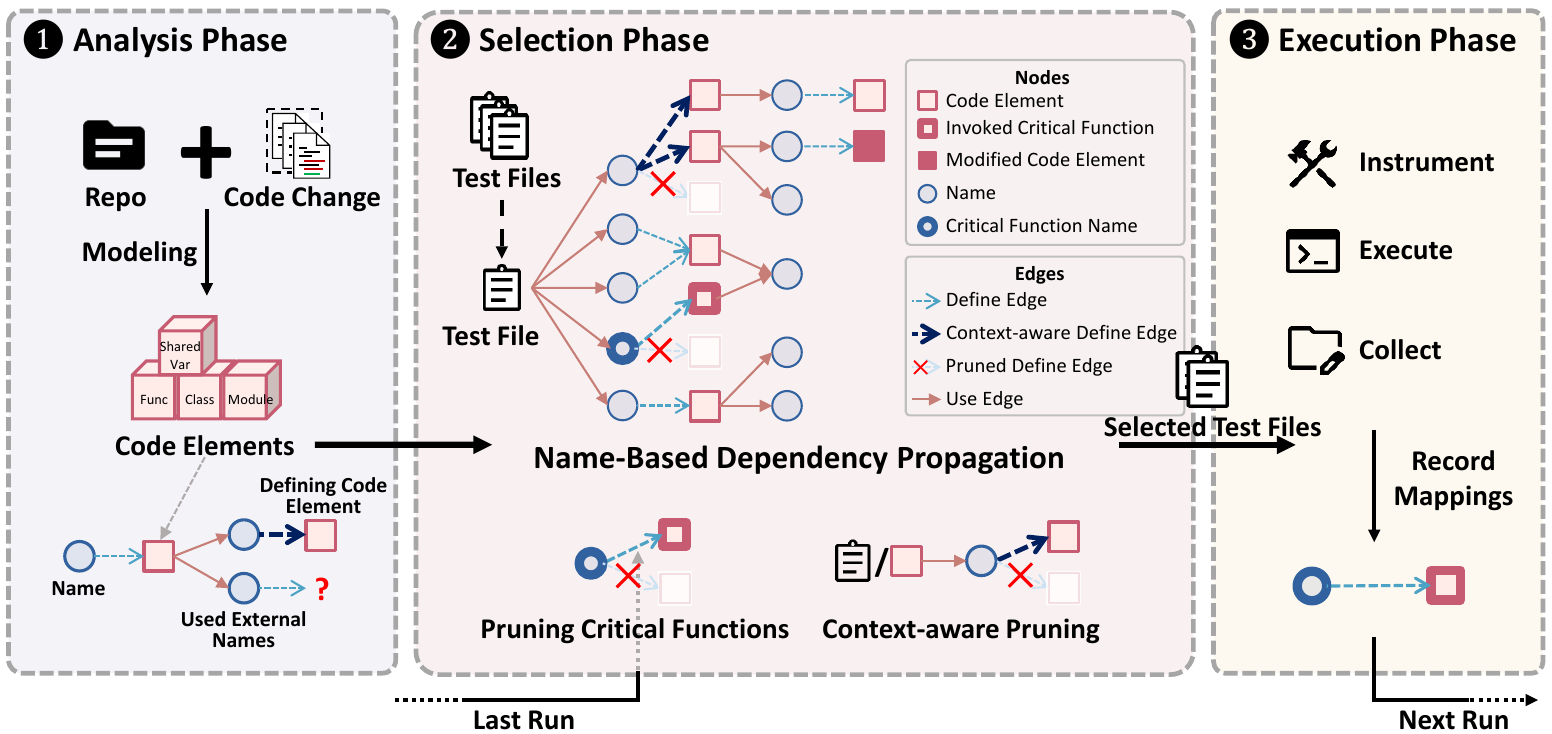}
    \caption{Overview of \appname.}
    \label{fig:overview}
\end{figure*}

\appname is a regression test selection technique for Python that determines which tests may be affected by a given code change using \emph{name-based dependency propagation}. 
For each test file, \appname identifies whether it may reach any modified code element by propagating dependencies through a lightweight name-element graph. 
As illustrated in Figure~\ref{fig:overview},  \appname operates in three phases: analysis, selection, and execution. 
The analysis phase extracts fine-grained code elements and their name dependencies, and maintains the metadata required for dependency propagation.
The selection phase performs propagation for each test file and applies pruning strategies to eliminate paths with no evidence of reachability. 
The execution phase runs the selected tests and records runtime metadata for use in subsequent runs.
Before applying \appname to a new commit, a one-time \emph{initialization run} is required. 
This run applies all three phases and executes the full test suite to establish the metadata needed for \emph{incremental runs} performed after every code change.

\subsection{Used Metadata}
\label{sec:metadata}

\appname maintains metadata defining the dependency information needed for 
\emph{name-based dependency propagation}.
These metadata are incrementally updated across runs to capture fine-grained dependencies between names and code elements. 

\paragraph{Code Elements.} In the analysis phase, the target project is parsed into four types of code elements: \texttt{Module}, \texttt{Class}, \texttt{Function}, and \texttt{SharedVariable}. 
Each code element is represented as 
\( 
    e = \langle \mathit{name}, \mathit{type}, \mathit{checksum}, \mathit{usedNames}, \mathit{file}, \mathit{optional(defClass)}\rangle 
\),
including its unqualified identifier ($\mathit{name}$), code element type ($type$), the hash of its bytecode ($\mathit{checksum}$), the set of external names it references ($\mathit{usedNames}$), and its defining file. For \texttt{Function} and \texttt{SharedVariable} elements, an additional attribute \texttt{defClass} specifies their defining class, if any.
\texttt{SharedVariable} includes global variables and class static variables, which are initialized at import time and serve as shared state.

\paragraph{Name-element graph.}
To formalize name-based dependency propagation, \appname models a bipartite name-element graph consisting of \emph{code element nodes} \( E \), one for each code element derived above; 
and \emph{name nodes} \( N \), 
which are all \textit{name} fields of code elements.
Directed edges are defined as two sets, i.e., \emph{definition edges} and \emph{usage edges}.
Definition edges \(
    Def = \{\, (n,e) \mid n \in N \land e \in E \land e.\mathit{name} = n \,\}
\) map each name to code elements defined under that name, indicating that resolving name \( n \) may yield code element \( e \). 
Usage edges \(
    Use = \{\, (e,n) \mid n \in N \land e \in E \land n \in e.\mathit{usedNames} \,\}
\) capture the external names referenced by each code element, 
indicating that \( e \) depends on the definition of \( n \).  
Propagation over these edges captures how names lead to code elements and how code elements introduce further name dependencies.

\paragraph{Import Graph.} 
The import graph is a mapping \(IG: s \mapsto \{\, s' \mid s' \text{ is imported in } s \,\}\)
from each source file $s$ to the set of files it directly imports. This information is used to resolve transitive module dependencies.

\paragraph{Accessed Names.} 
\appname maintains a mapping \(AN: t \mapsto \{\, n \mid n \in N \,\}\) that associates each test file \(t\) with the set of names it accesses at runtime, including those obtained indirectly (e.g., via reflection). 
This metadata captures implicit runtime dependencies not visible in static analysis.

\subsection{Analysis Phase}

The goal of the analysis phase is to construct the bipartite name-element graph described in Section~\ref{sec:metadata}. 
This includes deriving code elements, identifying the external names they reference, and establishing the definition and usage edges that connect code element nodes and name nodes. 
In addition, the analysis phase determines the set of code elements visible during test execution for each test file, and detects modified code elements for later use.

\paragraph{Constructing code elements.}
To extract the four types of code elements, \appname{} performs a source-level static analysis.

\noindent \textbf{Extracting used external names.}
The extraction process of used external names is shared across all four code element types.
\appname extracts these names by compiling the project and inspecting the Python bytecode.
For each snippet from which a code element is constructed, \appname identifies the corresponding bytecode range by mapping bytecode instructions to source line numbers and selecting those whose line numbers fall within the snippet.
It then reads identifier operands from the instructions flagged by \texttt{hasname}~\cite{pythonhasname} in Python’s \texttt{opcode} module, which reference entries in the \texttt{co\_names} table.
This bytecode-based approach avoids the need to trace variable definitions to determine whether an accessed identifier refers to a local variable or an external element, as local variables are stored separately in the \texttt{co\_varnames} table and are therefore naturally excluded \cite{pythondatamodel}.

\noindent \textbf{Constructing \texttt{Function} elements.}
The analysis constructs a \texttt{Function} code element for each top-level function or method definition and extracts its external names using the bytecode-based procedure described above.
Any function or class defined inside a function body is treated as part of that enclosing function and does not form a separate code element.

\noindent \textbf{Constructing \texttt{Class} elements.}
For each class definition, including its inheritance clause but excluding method definitions and static class variable definitions, the analysis constructs a \texttt{Class} code element.
The used external names of a \texttt{Class} element primarily include its superclass names, enabling \appname to determine when members inherited from those superclasses may be reachable.

\begin{wrapfigure}{r}{0.27\textwidth}
    \setlength{\abovecaptionskip}{0cm}
    \centering
    \includegraphics[width=0.6\linewidth]{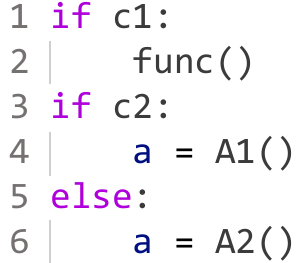}
    \caption{Example Code with \texttt{Module} and \texttt{SharedVariable} elements}
    \label{fig:analysis_code_example}
\end{wrapfigure}

\noindent \textbf{Constructing \texttt{Module} elements.}
The analysis constructs \texttt{Module} code elements to capture import-time behavior with side effects that may influence test execution.
For each module, the used external names are extracted from all top-level statements executed at import time (including compound constructs, such as \texttt{if}, \texttt{for}, or \texttt{try}) that contain function calls whose return values are not consumed. 
Such statements are treated as contributing to import-time side effects.
For example, in Figure~\ref{fig:analysis_code_example}, lines 1–2 form a \texttt{Module} element, yielding the external names $c1$ and $func$.

\noindent \textbf{Constructing \texttt{SharedVariable} elements.}
Global variables and class static variables are initialized at import time and constitute shared state that may be accessed across multiple contexts.
The analysis constructs a \texttt{SharedVariable} code element for each such variable, 
extracting used external names from top-level statements executed at import time that define or update that variable. 
For example, in Figure~\ref{fig:analysis_code_example}, lines 3–6 define the global variable $a$ and form its \texttt{SharedVariable} element, yielding the external names $c2$, $A1$, and $A2$.

Each constructed code element becomes a node in the name-element graph.
A definition edge conceptually connects the code element to the name under which it is defined, and its extracted external names give rise to usage edges to the corresponding name nodes.
The resulting conceptual graph forms the basis for name-based dependency propagation in the selection phase.

\paragraph{Determining visible code elements for each test file.}
Since a test can only reach code elements defined in files it imports, we use file-level visibility as an initial filter to eliminate provably irrelevant code elements.
\appname constructs the import graph \(IG\) by statically analyzing import statements in source files to extract file-level import relationships.
To handle Python’s import semantics, when a submodule is imported, \appname also records the \texttt{\_\_init\_\_.py} files of all its parent packages as imported files.
Given this graph, \appname computes, for each test file, the set of source files it transitively depends on by performing a graph traversal starting from the test file.
From this set, \appname derives \(Visible[t]\), the set of code elements whose defining files are reachable from $t$ in the import graph and are therefore visible during the execution of $t$.

\paragraph{Detecting modified code elements.}
To detect modified code elements, \appname computes a checksum for each code element from its Python bytecode and compares it against the previously stored checksum. 
Before computing the checksum, \appname removes or normalizes non-semantic information in the bytecode that does not reflect actual code changes, such as virtual memory addresses or other compilation-specific metadata. 
Since comments are not preserved in Python bytecode, they naturally do not affect the computed checksums. 
The modified set $M$ contains all code elements whose checksums have changed or are not present in the previous run.

\smallskip
The analysis phase runs during both the initialization run and the incremental runs.
In the initialization run, it analyzes all source files to construct code elements and their metadata, which are cached.
In the incremental runs, it loads the cached results, re-analyzes only modified files, and updates the cache accordingly.
Modified code elements are identified only during subsequent runs.

\subsection{Selection Phase}

\begin{figure*}[t]
\centering

\begin{minipage}[t]{0.60\textwidth}
\fontsize{8pt}{9pt}\selectfont
\begin{algorithm}[H]
\SetAlgoHangIndent{0pt}
\SetInd{0.4em}{0.3em}
\DontPrintSemicolon
\caption{\fontsize{8pt}{9pt}\selectfont \mbox{Name-based Dependency Propagation}}
\label{alg:namebdp}
\KwIn{
$\mathcal{T}$: set of all test files; \\
$AN$: Accessed Names metadata; \\
$Def$: set of definition edges $(n,e)$; \\
$Use$: set of usage edges $(e,n)$; \\
$M$: set of modified code elements; \\
$Visible$: mapping $t \mapsto$ visible code elements.
}
\KwOut{
$\mathcal{T}'$: set of affected test files.
}
$\mathcal{T}' \leftarrow \emptyset$ \;
\ForEach{$t \in \mathcal{T}$}{
    $\textit{reachable\_names} \leftarrow$ external names used in $t$ \;
    $Modules_t \leftarrow \{\, m \in Visible[t] \mid m.type = \textsc{Module} \}$ \;
    \ForEach{$m \in Modules_t$}{
        $\textit{reachable\_names} \mathrel{\cup\!=} m.usedNames$
    }
    $\textit{reachable\_names} \mathrel{\cup\!=} AN[t]$ \;
    $\textit{reachable\_elements} \leftarrow \emptyset$ \;
    $\textit{work\_stack} \leftarrow \textit{reachable\_names}$ \;
    \Repeat{no new names are added}{
        \While{$\textit{work\_stack}$ not empty}{
            $n \leftarrow \textit{pop(work\_stack)}$ \;
            $E \leftarrow \{\, e \mid (n,e)\in Def \land e \in Visible[t] \,\}$ \;
            $E \leftarrow \{\, e \in E \mid e.\textit{class} = \textit{null} \;\lor\; e.\textit{class} \in \textit{reachable\_names} \,\}$ \;
            prune $E$ with Algorithm~\ref{alg:prune} \;

            $\textit{reachable\_elements} \mathrel{\cup\!=} E$ \;
            \ForEach{$e \in E$}{
                $names \leftarrow \{\, n' \mid (e,n') \in Use \,\}$ \;
                add new $names$ to $\textit{reachable\_names}$ and $\textit{work\_stack}$ \;
            }
        }
        refill $\textit{work\_stack}$ with names of \texttt{Function} and \texttt{SharedVariable} elements defined in classes \;
    }
    \If{$\exists e \in \textit{reachable\_elements}$ s.t. $e \in M$}{
        $\mathcal{T}' \leftarrow \mathcal{T}' \cup \{t\}$ \;
    }
}
\Return{$\mathcal{T}'$}
\end{algorithm}
\end{minipage}
\hfill
\begin{minipage}[t]{0.39\textwidth}
\fontsize{8pt}{9pt}\selectfont
\begin{algorithm}[H]
\SetAlgoHangIndent{0pt}
\SetInd{0.4em}{0.3em}
\DontPrintSemicolon
\caption{\fontsize{8pt}{9pt}\selectfont \mbox{Code Element Pruning}}
\label{alg:prune}
\KwIn{
$n$: name being propagated; \\
$type(n)$: name type (\textit{non-attr}, \textit{sure-attr}, or \textit{amb-attr}); \\
$ctx$: optional context (defining file or class); \\
$E$: candidate code element nodes associated with $n$; \\
$IC$: Invoked Critical Functions; \\
$\textit{reachable\_names}$: current set of reachable name nodes.
}
\KwOut{
$E$: pruned set of code element nodes.
}
\If{$n$ corresponds to a critical function}{
    $\textit{E} \mathrel{\cap\!=} IC[t]$ \;
}
\If{$type(n) = \textsc{non-attr}$}{
    \If{definition of $n$ exists in $ctx.file$}{
        \Return{definition of $n$} \;
    }
    \Else{
        $def\_n \leftarrow$ explicit import target of $n$ \;
        \Return{$def\_n$ if exists, otherwise $E$} \;
    }
}
\If{$type(n) = \textsc{sure-attr}$}{
    \Return{$\{\, e \in E \mid e.class \in \text{hierarchy}(ctx.class) \,\}$} \;
}
\Return{$E$}
\end{algorithm}
\end{minipage}
\end{figure*}

In the selection phase, \appname performs name-based dependency propagation for each test file with respect to the current code change, realized as a reachability computation over the bipartite name-element graph described in Section~\ref{sec:metadata}. 
A test file is selected for re-execution if it can reach any modified code element node. 
Algorithm~\ref{alg:namebdp} outlines the overall procedure.

For each test file $t$, \appname first initializes the set \textit{reachable\_names} with all external names referenced by code elements defined in $t$ (line 3). 
It then incorporates import-time behavior by adding the external names referenced by the \texttt{Module} elements visible to $t$ (line 4–6).
The set is further augmented using \textit{Accessed Names} metadata, which records names implicitly accessed during previous executions of $t$ (line 7).
These names form the initial frontier of reachable name nodes.

After initialization, \appname iteratively expands \textit{reachable\_names} until a fixed point is reached (lines 8-20).
A working stack, \textit{work\_stack}, is initialized with all current name nodes in \textit{reachable\_names} (line 9). 
For each name node $n$ popped from the stack, \appname follows definition edges by collecting all code element nodes $e$ such that $(n,e)\in Def$ and $e$ is visible to $t$ (line 12-13).
The resulting candidates are then pruned.
Specifically, \appname filters out code elements whose defining class is not present in the \textit{reachable\_names} set (line 14).
A class outside \textit{reachable\_names} is considered inaccessible, implying that its attributes are not reachable.
The remaining code element nodes are recorded as reachable (line 16).
For each recorded code element node $e$, \appname follows usage edges by adding newly reached name nodes \(n'\) such that $(e,n')\in Use$ to both \textit{reachable\_names} and \textit{work\_stack} (line 17–19).
This alternation of traversing definition and usage edges continues until \textit{work\_stack} becomes empty. 
Because new class names may appear during propagation, \appname then refills the \textit{work\_stack} with the names of \texttt{Function} and \texttt{SharedVariable} elements defined in classes to ensure that attribute-related code elements depending on newly reachable classes are not prematurely excluded (line 20).
If no new names are added during this iteration, propagation has converged.
Finally, if any recorded reachable code element node belongs to the modified set $M$, the corresponding test file is marked as affected and selected for execution (lines 22–23).

\begin{figure*}[t]
\centering

\begin{subfigure}{0.54\textwidth}
    \centering
    \includegraphics[width=\linewidth]{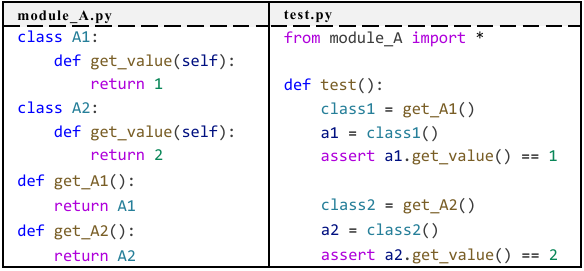}
    \caption{Example project.}
    \label{fig:selection_example_code}
\end{subfigure}
\hfill
\begin{subfigure}{0.45\textwidth}
    \centering
    \includegraphics[width=\linewidth]{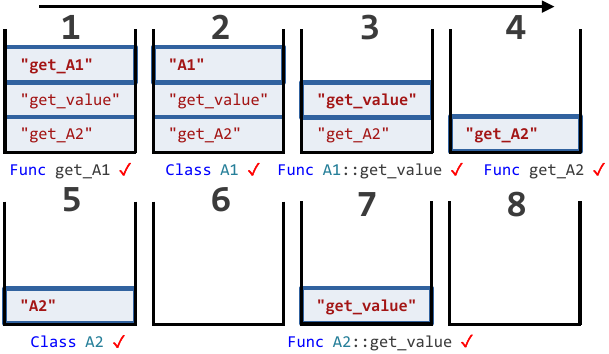}
    \caption{Evolution of the \textit{work\_stack}.}
    \label{fig:selection_example_stack}
\end{subfigure}

\caption{Example of name-based dependency propagation.}
\label{fig:selection_example}
\end{figure*}

Figure \ref{fig:selection_example} illustrates the propagation procedure using an example.
Figure \ref{fig:selection_example_code} shows the example project, while Figure \ref{fig:selection_example_stack} depicts the evolution of the \textit{work\_stack}, which maintains the frontier of name nodes to be processed.
Initially (state 1), the \textit{work\_stack} contains the external names referenced by \texttt{test.py}, namely \texttt{get\_A1}, \texttt{get\_value}, and \texttt{get\_A2}.
When \texttt{get\_A1} is popped, \appname resolves it to the function \texttt{get\_A1}, whose body references the class name \texttt{A1}.
Thus \texttt{A1} is pushed to the stack (state 1-2).
Subsequent steps proceed similarly.
Notably, when name \texttt{get\_value} is processed in state 3-4, only \texttt{A1::get\_value} is considered reachable, since class name \texttt{A2} is not yet accessible and \texttt{A2}'s attributes are therefore pruned.
After \texttt{A2} becomes reachable and the stack is exhausted (state 6), \appname refills the \textit{work\_stack} with attribute names, causing name \texttt{get\_value} to be reconsidered and \texttt{A2::get\_value} to be correctly included (states 7–8).    

\subsection{Execution Phase}
\label{sec:exec}

In the execution phase, \appname runs test files and collects runtime metadata to support future test selection.
During the initialization run, the full test suite is executed, whereas in incremental runs only the selected tests are executed.

\paragraph{Augmenting Accessed Names via Dynamic Analysis.}
\appname assumes that if the Python interpreter can access a name during test execution, its definition may be used. 
However, not all uses appear as direct name accesses, which might cause the approach to miss some dependencies. 
We address two types of indirect name accesses by dynamically analyzing the executing regression tests: operator overloading and reflection.
For operator overloading, rather than matching every possible overloaded operator, \appname records all operator-related names (e.g., \texttt{\_\_add\_\_}) under a special key "\textit{*}" in a map $AN$ 
(for ``accessed names''), meaning they are shared by all test files.
For reflection, \appname focuses on introspection, which inspects program structure or object attributes at runtime, e.g., \texttt{builtins.getattr}. 
The approach monkey-patches common built-in introspection functions to identify accessed names based on their parameters or return values, and adds these names to $AN[t]$ for the currently executing test file.
Names accessed during module import are likewise recorded under the shared key "\textit{*}", so they are treated as reachable to all test files.
Dynamic code execution through \texttt{eval} or similar mechanisms is not tracked, as it is rare and prohibitively expensive to analyze safely \cite{li2019understanding,aakerblom2014tracing}.

\paragraph{Dynamic Import Events.}
While standard \texttt{import} statements are analyzed in the analysis phase to construct the import graph \(IG\), imports performed through the \texttt{importlib} API require dynamic handling, as the module name is often stored in a variable whose value may be constructed at runtime and is hard to resolve statically.
\appname intercepts imports triggered through the \texttt{importlib} API by monkey-patching \texttt{importlib.import\_module}, and augments \(IG\) with the captured dynamic import events.
These dynamically observed imports are recorded and incorporated into \(IG\) in the analysis phase during all future incremental runs.

\subsection{Pruning Mechanisms}

The approach explained so far conservatively assumes that any definition with a specific name is a dependency.
In practice, 
this conservative approach often causes the set of reachable code elements to grow far beyond what a test can actually reach. 
A single name node may pull in large portions of the graph, which recursively trigger further expansions. 
This dependency cascade inflates the reachable code element set and reduces the precision of selection.
To control this effect, \appname integrates two pruning mechanisms into the dependency propagation procedure. 
Both refine the candidate set of code element nodes that are considered valid successors of a name node and discard paths that have no evidence of being reachable in a given test.

\paragraph{Pruning critical functions.}
Some name nodes correspond to many code element nodes.
For example, a single method name may correspond to method implementations in many different classes.
When propagation reaches such a name node, it may induce a large dependency cascade that is unlikely to reflect the behavior of an individual test.
To mitigate this effect, \appname identifies functions whose name nodes possibly lead to many other nodes and requires runtime evidence before the propagation is allowed to proceed through them.

Critical functions are identified during the initialization run, where \appname performs dependency propagation for all tests. 
For each name node, \appname aggregates outgoing name nodes reachable through all of its candidate code elements. 
Names with the largest aggregated expansions are primary sources of dependency cascades. 
\appname selects the top \(N\) such names as \emph{critical function names},  where $N$ is a configurable parameter.
All functions defined under these names are treated as \emph{critical}.
In the initialization and subsequent runs, before executing tests, \appname inserts lightweight instrumentation at the entry points of all critical functions.
Each instrumentation probe records the first invocation of the function during the execution of a test file, and ignores subsequent invocations, ensuring negligible runtime overhead.
If a critical function is called during module import, the probe records the event under the shared key ``\(*\)''.
In the selection phase, when propagation encounters a critical function, \appname prunes the corresponding code element node unless an invocation was observed for the current test (line 1-2 in Algorithm~\ref{alg:prune}).
We preserve two exceptions and never prune these functions even without runtime evidence: newly added functions and methods whose overriding implementations in subclasses were removed, since both can introduce new invocations that were not observed in prior executions.

\paragraph{Pruning with context-aware name-element matching}
The second pruning mechanism refines dependency propagation by exploiting the context in which names are used. 
For many used external names, contextual information enables more precise name-element matching than trivial name-based matching, which prevents propagation to spurious defining code element nodes.

During the analysis phase, \appname classifies each used external name into one of three categories based on its bytecode-level access pattern: 
\begin{itemize}[topsep=0pt, itemsep=1pt]
\item \emph{non-attr:} names accessed without a dot operation (i.e., not via \texttt{LOAD\_ATTR}). These are typically explicitly imported or defined in the current module, making definitions straightforward to resolve.
\item \emph{sure-attr:} attribute accesses on \texttt{self} or \texttt{cls}. Their definitions can be reliably resolved within the enclosing class, its superclasses, or its subclasses. 
\item \emph{amb-attr:} attribute accesses on objects other than \texttt{self} or \texttt{cls}. 
These may refer to attributes of arbitrary objects, including modules, and are therefore \textit{ambiguous} in origin.
\end{itemize}

In the selection phase, when a name node becomes reachable, \appname records the context of the code element node that produced it, including its defining file $ctx.file$ and, if any, its defining class $ctx.class$.  
Pruning proceeds according to the name category. 
For \emph{non-attr} names, \appname attempts to locate its definition within the same file $ctx.file$ (Algorithm~\ref{alg:prune}, lines 5-6). 
If none is found, it traces explicit import statements to locate the imported definition.  
When such a definition is found, only that code element node is retained. 
Otherwise, 
all candidate elements are kept to ensure safety (Algorithm~\ref{alg:prune}, lines 8–9).
For \emph{sure-attr} names, \appname retains code element nodes whose defining classes lie in the hierarchy of the referencing class $ctx.class$ (Algorithm~\ref{alg:prune}, line 11).  
These matching steps can be precomputed to avoid repeated lookups during propagation.
For \emph{amb-attr} names, \appname preserves all candidate code element nodes to maintain safety.

\subsection{Special Handling of Decorators}

Decorators in Python are functions executed at definition time, wrapping a class or function and binding the returned object to the original name \cite{pythondecorator}. 
Since decorators are typically executed during module import, they need to be handled with care. 
\appname distinguishes between two kinds of decorators.
 \emph{Functional decorators} modify or extend the behavior of the decorated object. They are treated as external names used by the decorated function or class, since their effects manifest only when the decorated object is invoked.
 \emph{Registry decorators}, by contrast, register the decorated object for later use.
\appname identifies such decorators heuristically by matching configurable keywords commonly used in registration patterns (e.g., \texttt{register}, \texttt{router}).
Because such decorators determine when and how the registered objects are invoked, their behavior is difficult to fully capture through static analysis. 
To remain safe, for classes decorated in this way, \appname conservatively adds them to $AN[*]$. 
For functions, \appname instruments them and monitors their execution during test runs.
If a decorated function is executed by a test $t$, it is added to $AN[t]$. 
Finally, if a registry decorator itself is modified or newly introduced, all test files are marked as affected, as the registration behavior of the system may have changed.

%% file: sections/evaluation.tex
\section{Evaluation}
To evaluate \appname, we investigate the following research questions: 

\begin{itemize}[topsep=0pt, itemsep=1pt]
    \item \textbf{RQ1:} How effective is \appname in reducing tests and testing time while remaining safe?
    \item \textbf{RQ2:} What is the computational overhead introduced by \appname?
    \item \textbf{RQ3:} How much do the two pruning mechanisms contribute to the effectiveness?
    \item \textbf{RQ4:} How does the choice of the parameter \textit{N} for identifying critical functions affect effectiveness?
\end{itemize}

\subsection{Experimental Setup}
\label{sec:evalsetup}

\paragraph{Dataset construction.}
Python projects can grow to hundreds of thousands of lines of code, yet prior RTS studies for Python consider only projects up to 60k lines~\cite{kauhanen2021regression,maurina2025babelrts}.
To provide a more comprehensive assessment of \appname, we curate a new dataset of 10 open-source Python projects.
We first define the following selection criteria: 
(i) Python is the primary language of the project; 
(ii) the project is compatible with Pytest; 
(iii) executing the full test suite requires more than 30 seconds; and
(iv) the project contains more than 20 test files.
Network-dependent projects are excluded because their test execution times tend to be unstable.
Based on these criteria, 
we first consider the projects used by the SWE-bench benchmark~\cite{jimenez2023swe,wang2025solved}, which represent widely used, actively maintained Python systems, and identify seven projects that satisfy all criteria. 
Then, to increase diversity, we randomly sample additional projects from GitHub repositories with more than 1,000 stars and more than 500 commits, retaining the first three sampled projects that meet the same criteria.
These projects are summarized in Table~\ref{tab:dataset}. 
They contain, on average, 130k lines of Python code, with sizes ranging from 12k to 426k. 
For each project, we collect 50 consecutive commits, following the setup used in prior work \cite{zhang2024hybrid}. 
Each commit must build successfully, as a commit that fails to build cannot execute any tests.
We also require that each commit modifies at least one Python source file. 
Commits that only touch test files or non-Python files are filtered out, ensuring that our evaluation focuses on source-code changes relevant to RTS. 
The complete list of selected commit hashes is available in our replication package~\cite{package}.

\input{tables/dataset}

\paragraph{Ground truth extraction.}
To compute a ground truth for test selection, we instrument every modified function and execute each test file to determine whether the instrumented functions are invoked.
For modified global variables and class static variables, we use a language server to trace their references until functions are reached and then instrument these functions. 
A language server is suitable here because ground truth extraction requires only a one-time, precise resolution of actual usages. 
Although individual language server queries are fast, applying such analysis in RTS would require resolving call targets or references at a large number of program locations on every commit, resulting in prohibitive cumulative overhead.
Therefore, we use the language server only for offline ground truth collection.
One complication is that a function may execute during module import for initialization. 
Because of Python's eager importing feature, such functions often run even when the test never uses their initialized state. 
To avoid such false positives, we manually inspect invocations that occur only during import and not during test execution, and discard test files that do not actually use the initialized states.
To the best of our knowledge, this is the first Python RTS dataset that provides a ground truth.

\paragraph{Used metrics.}
RTS evaluation focuses on effectiveness and safety. 
Following prior work \cite{maurina2025babelrts}, effectiveness is evaluated along two dimensions. \emph{Test reduction} measures how many test files are skipped on average, 
computed as \(\text{TestR} = (T_{\text{all}} - T_{\text{sel}}) / T_{\text{all}}\), 
where $T_{\text{all}}$ and $T_{\text{sel}}$ denote the total and selected test files across all commits.
As a complementary metric to test reduction, we also measure \emph{precision}, which quantifies how many selected test files are actually affected according to the ground truth, computed as \(\text{Precision} = T_{\text{aff}} / T_{\text{sel}}\), where $T_{\text{aff}}$ denotes the number of selected test files that are affected.
Moreover, \emph{time reduction} measures savings in end-to-end testing time, including initialization (if any), analysis, selection, and test execution, 
computed as \(\text{TimeR} = (t_{\text{orig}} - t_{\text{rts}}) / t_{\text{orig}}\), 
where $t_{\text{orig}}$ and $t_{\text{rts}}$ denote the cumulative testing time without and with RTS.
Safety is evaluated using \emph{safe rate}, which measures the fraction of commits for which all tests that are affected according to the ground truth are selected, 
computed as \(\text{SafeR} = C_{\text{safe}} / C_{\text{total}}\), 
where $C_{\text{safe}}$ and $C_{\text{total}}$ denote the number of safe commits and the total number of commits.

\paragraph{Implementation details.}
Pytest fixtures are injected into test functions through parameters, which are treated as local variables and therefore do not appear in $\mathit{usedNames}$. 
This would cause \appname to miss dependencies introduced by fixtures. 
To avoid this, we add the names of injected fixtures to the $\mathit{usedNames}$ of the receiving functions so they participate in dependency propagation.
\appname monitors the invocation of critical functions during test execution, while we observe that shared global state may cause missing invocations. 
For example, one test may initialize a global variable, preventing subsequent tests from invoking the same initialization code. 
To avoid such interference, we implement isolated test execution atop \texttt{pytest-xdist}: 
each test file runs in a separate process while preserving overall serial order. 
This yields consistent runtime metadata and safe dependency tracking.
One project, \textit{pylint}, is incompatible with \texttt{pytest-xdist}, so isolation is disabled for it. 
Our results show that disabling isolation for \textit{pylint} does not lead to substantial safety loss. 
Isolation is enabled only for \appname and disabled for all baselines.
Unless otherwise noted, the parameter $N$ for identifying critical functions is set to 500.

\subsection{RQ1: Effectiveness and Safety}

\paragraph{Approach.}
We evaluate \appname against two baselines: BabelRTS~\cite{maurina2025babelrts} and EkstaP. 
Existing RTS techniques can be categorized into file-level and function-level approaches \cite{yoo2012regression}.
BabelRTS is a file-level RTS technique and represents the state of the art for Python.
It relies on static file-level dependencies but ignores implicit imports that may affect program behavior, 
which leads to unsafe test selection.
To compare \appname with a safe file-level RTS, we implement an additional baseline, \emph{EkstaP}, following the file-level RTS paradigm exemplified by Ekstazi~\cite{gligoric2015practical} for Java.
EkstaP reuses the import graph infrastructure of \appname and accounts for implicit parent-package imports and dynamic imports via \texttt{importlib}, as described in Section~\ref{sec:exec}.
Rather than faithfully reproducing Ekstazi in Python, EkstaP is designed to provide a strong and safe file-level RTS baseline with as much reduction as possible while preserving safety.
We do not include pytest-rts~\cite{kauhanen2021regression}, a coverage-based RTS technique that selects test functions based on previously observed coverage, because our ground truth and safety metric are defined at the test-file level, preventing a fair comparison. 
Moreover, prior work~\cite{maurina2025babelrts} shows that BabelRTS achieves substantially better performance than pytest-rts, so we focus on BabelRTS as a stronger baseline.
At present, there is no function-level RTS technique specifically designed for Python.
Function-level RTS techniques developed for other languages rely on call graph construction.
In Python, call graph analysis suffers from scalability and precision limitations \cite{bouzenia2024dypybench}, 
making porting such techniques non-trivial and unreliable.

\paragraph{Results.}

\input{tables/rq1}

Table~\ref{tab:rq1} reports the performance of \appname compared with the baselines. 
On average, \appname skips 69.90\% of test files and reduces end-to-end testing time by 45.59\%, achieving 86.63\% of the maximum test reduction indicated by the ground truth (69.90\% out of 80.69\%), while attaining a precision of 66.65\%, substantially higher than EkstaP (20.77\%) and BabelRTS (24.25\%). 
Although the precision is not perfect, only 19.31\% of test files are affected by a typical code change on average according to the ground truth, allowing \appname to still achieve test reduction close to the ground truth.
\appname's test reductions hold across projects of very different scales.
For example, it eliminates 73.51\% of test files in \textit{sympy} (426k LoC) and 94.33\% in \textit{pvlib} (28k LoC). 
Several projects show more modest reductions, yet \appname still removes more than half of their tests.
A key factor affecting the effectiveness of \appname on these projects is that certain critical functions execute during module import. 
For such functions, \appname must conservatively assume that any test importing the defining file may access the functions. 
This effect is most visible when changes occur inside global code that includes side-effecting calls. 
In \textit{matplotlib}, for instance, five out of fifty commits modify the function \texttt{\_get\_executable\_info}, which is executed when \texttt{matplotlib/testing/compare.py} is imported. 
For these commits, \appname eventually selects all tests. 

Regarding safety, \appname is safe for 99.6\% of all commits, i.e., it is unsafe for only two commits.
One unsafe commit appears in \textit{pylint}\footnote{https://github.com/pylint-dev/pylint/commit/7d6f3f230e038eabee2efc75628329fe74aa943e}, where two test files are missed. 
This is because shared state left in the global \texttt{astroid.MANAGER} cache by an earlier test causes the resolution logic to short-circuit and never reach the critical function \texttt{AggregationsHandler.handle}. 
The code element node is incorrectly pruned, preventing dependency propagation from reaching the modified function. 
This test interference occurs only because our isolator cannot be applied to \textit{pylint}, as discussed in Section~\ref{sec:evalsetup}.
\appname also misses one test in \textit{sphinx}\footnote{https://github.com/sphinx-doc/sphinx/commit/c76c2ad63c172e5ebab8bdb965286cfde5ca4491}. 
Specifically, \textit{sphinx} and \textit{docutils} interact through a callback protocol, in which \textit{docutils} explicitly invokes \texttt{dispatch\_visit} defined in \textit{sphinx}.
Because this invocation is hard-coded in the external library, i.e., \textit{docutils}, and \appname does not analyze library code, the callback cannot be inferred.

BabelRTS skips 28.36\% of test files and yields a 21.95\% reduction in end-to-end testing time.
Its safety, however, is inconsistent. 
For 6 of the 10 projects, the Safe Rate falls below 90\%.
The most common safety issues stem from BabelRTS ignoring implicit parent package imports.
The safety degradation is most pronounced in \textit{matplotlib} and \textit{pylint}, where BabelRTS achieves high test reduction (97.63\% and 60.84\%) but attains safe rates of only 14.00\% and 48.00\%.
In \textit{matplotlib}, the unsafety is caused by BabelRTS assuming that package sources are located under the project root, but \textit{matplotlib} places its code under \texttt{lib/matplotlib}, leading to missed dependencies.
In \textit{pylint}, unsafety instead arises from extensive runtime plugin loading via \textit{importlib}, which BabelRTS’s static-only analysis cannot capture.
EkstaP, which incorporates these missing dependency edges and handles implicit and dynamic imports safely, achieves 7.42\% test reduction and 3.10\% time reduction.
For 3 of the 10 projects, EkstaP cannot skip any test. 
By analyzing EkstaP’s import graph, we find that Python’s eager importing causes severe inflation of file-level dependencies: on average, a single test file imports 83.27\% of source files, and 80.92\% of source files are imported by every test file.
As a result, file-level RTS leaves little opportunity for meaningful test reduction, and the overhead of EkstaP’s required initialization run to capture dynamic imports can outweigh its benefits, leading to limited or even negative time reduction.
These results highlight that while file-level RTS performs well in languages like Java \cite{gligoric2015practical,legunsen2017starts}, it is substantially less effective in Python.

\vskip 1mm
\noindent \fbox{
	\parbox{0.95\linewidth}{\textbf{Answer to RQ1}: \appname skips 69.90\% of test files, reduces testing time by 45.59\%, and is safe on 99.6\% of commits. BabelRTS is often unsafe, while EkstaP remains safe but reduces only 7.42\% of tests and 3.10\% of time. Compared with this safe, file-level baseline, \appname provides an order of magnitude higher test and time reductions.}
}

\subsection{RQ2: Computational Overhead}

\paragraph{Approach.}
\appname introduces overhead during analysis, selection, and execution. 
RQ2 quantifies this overhead by breaking it down into three components: 
\emph{initialization overhead}, which refers to the one-time cost of the initialization run; 
\emph{selection overhead}, which measures the cost of identifying affected tests on each commit and covers the analysis and selection phases; 
and \emph{runtime overhead}, which captures the additional execution time introduced by instrumentation, such as probes inserted into critical functions.
Together with actual test execution time, these components form the end-to-end testing cost when \appname is enabled.
To measure the runtime overhead introduced by \appname, we record test execution time with and without instrumentation for the same selected tests, and take their difference as the runtime overhead.

\paragraph{Results.}

\input{tables/rq2}

Table~\ref{tab:rq2} reports the overhead introduced by \appname. 
The \textit{RunAll} column shows the time required to execute the full test suite across 50 commits. 
Under the \textit{Time} column, the \textit{Test}, \textit{Runtime}, \textit{Select}, and \textit{Init} subcolumns denote the actual test execution time and the runtime, selection, and initialization overhead, each as a percentage of \textit{RunAll}. 
The \textit{Selected} column reports the percentage of tests selected by \appname. 
On average, \appname introduces 10.04\% overhead relative to RunAll, consisting of 4.45\% runtime overhead, 3.22\% selection overhead, and 2.38\% initialization overhead. 
Initialization overhead remains stable at roughly 2\% across projects. 
This is expected because initialization includes one full test-suite execution, while RunAll includes fifty.
For relatively simpler projects, such as \textit{seaborn} and \textit{loguru}, the cost of static analysis is negligible, so initialization overhead is dominated by the test execution. 
Selection overhead varies with project scale and the relative cost of test execution.
Large and structurally complex projects incur higher selection overhead. 
For instance, \textit{sympy}, the largest project in our dataset with 426k lines of code, has one of the highest selection overheads at 7.43\%. 
\textit{sphinx} and \textit{xarray} have similar code sizes (106k and 112k LoC), yet \textit{sphinx} has far faster tests (6647.6s vs.\ 36311.8s), making its selection overhead proportionally larger (8.24\% vs.\ 1.04\%).
Runtime overhead originates primarily from instrumentation, in particular the probes inserted into critical functions. 
Projects whose tests invoke certain functions intensively incur higher runtime overhead. 
For example, \textit{sympy} and \textit{xarray} are computation-heavy mathematical libraries and show runtime overheads of 10.64\% and 5.72\%. 
In contrast, for projects dominated by lightweight operations, runtime overhead is negligible. 
For instance, \textit{seaborn} shows only 0.2\% runtime overhead.

To understand at what point in time enabling \appname pays off, Figure~\ref{fig:rq2_cumulative} shows the \emph{cumulative relative testing time}, i.e., the ratio between testing time incurred up to a given commit under \appname and the time required to run all tests for the same set of commits.
We present four representative projects: two where \appname achieves strong reductions (\textit{seaborn} and \textit{sklearn}) and two where the gains are comparatively smaller (\textit{sympy} and \textit{sphinx}).
The \textit{Average} curve denotes the mean across all ten projects.
Overall, cumulative relative testing time falls below 1.0 within five commits on average, and within ten commits even for projects where \appname provides more modest benefits, meaning that \appname’s benefits outweigh its initialization overhead quickly.

\vskip 1mm
\noindent \fbox{
	\parbox{0.95\linewidth}{\textbf{Answer to RQ2}: \appname introduces an average overhead accounting for 10.04\% of the full test-suite execution time. The overhead consists of 2.38\% initialization overhead, 3.22\% selection overhead, and 4.45\% runtime overhead. Enabling \appname pays off after five commits, on average.}
}

\subsection{RQ3: Impact of Pruning Mechanisms}

\paragraph{Approach.} 
\appname incorporates two pruning mechanisms to filter unreachable code element nodes and avoid dependency cascades: 
critical-function pruning (CF) and context-aware name-element matching (NEM). 
To evaluate their contributions to effectiveness, we run three variants of \appname by disabling each mechanism individually (w/o CF, w/o NEM) and both together (w/o CF\&NEM), and then compare their performance with that of \appname.

\input{tables/rq3}

\paragraph{Results.} 
Table~\ref{tab:rq3} summarizes the results. 
Pruning critical functions has the largest impact. 
Removing CF reduces test reduction by 30.67\% and time reduction by 33.17\%, relatively. 
This confirms that our strategy for identifying critical functions enables \appname to more accurately identify the code elements that are truly affected.
Removing NEM causes decreases of 2.24\% and 6.79\% in test and time reduction, respectively. 
NEM primarily improves time reduction by removing code element nodes that are not reachable, thus preventing propagation from exploring their subsequent paths and thereby reducing selection overhead.
This benefit appears most noticeably in projects with relatively time-consuming selection phases.
For example, in \textit{sympy}, where selection accounts for 7.43\% of the RunAll time, NEM reduces the selection overhead by 36.48\%, which improves overall time reduction by an additional 4.27 percentage points.
In \textit{dask}, removing NEM reduces the time reduction by 27.65\%, because NEM not only reduces the selection overhead but also helps \appname avoid the tests that are relatively slow. 
When both pruning mechanisms are removed, test reduction and time reduction drop by 36.46\% and 46.75\%, respectively. 
Even without any pruning, \appname still reduces more tests and time than EkstaP and BabelRTS, 
thanks to its fine-grained dependency analysis that exploits information unavailable to file-level RTS.

\vskip 1mm
\noindent \fbox{
	\parbox{0.95\linewidth}{\textbf{Answer to RQ3}: Removing critical-function pruning causes the largest decline, reducing test and time reduction by 30.67\% and 33.17\%. Removing context-aware name-element matching leads to smaller but still substantial declines. Disabling both mechanisms causes the largest drop, showing that the two pruning mechanisms are essential for \appname.}
}

\subsection{RQ4: Sensitivity to the Parameter N}

\paragraph{Approach.} 
\appname identifies the top-$N$ critical function names during the initialization run and monitors the invocation of the corresponding functions during subsequent test execution.
To study how the choice of $N$ affects the effectiveness of \appname, we conduct a sensitivity analysis over different values of $N$.
We evaluate \appname with $N$ set to 0, 100, 300, 500, 700, and 900.

\paragraph{Results.} 

\begin{figure*}[t]
\centering

\begin{subfigure}{0.49\textwidth}
    \centering
    \includegraphics[width=\linewidth]{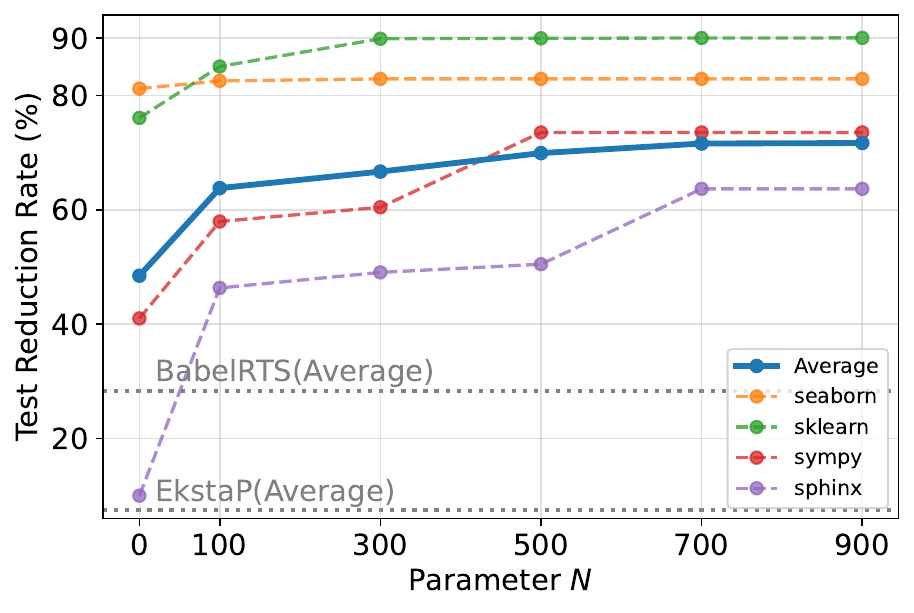}
    \vskip -2mm
    \caption{Test reduction.}
    \label{fig:rq4_testR}
\end{subfigure}
\hfill
\begin{subfigure}{0.49\textwidth}
    \centering
    \includegraphics[width=\linewidth]{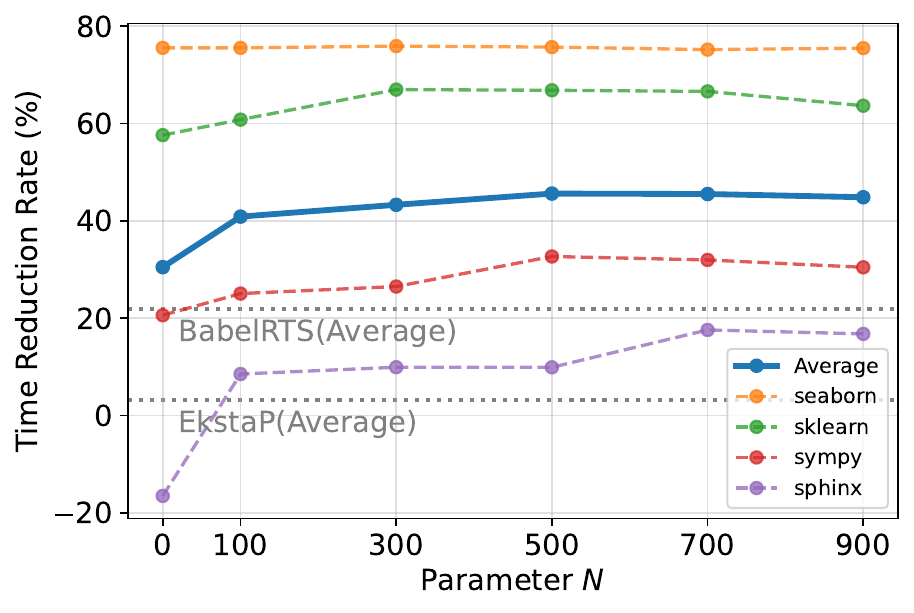}
    \vskip -2mm
    \caption{Time reduction.}
    \label{fig:rq4_timeR}
\end{subfigure}
\vskip -3mm
\caption{Parameter sensitivity analysis (RQ4).}
\label{fig:rq4}
\end{figure*}

Figure~\ref{fig:rq4} summarizes the results for four representative projects, chosen to cover both strong and relatively weaker effectiveness of \appname.
The reported average is computed over all ten projects.
Overall, test reduction increases monotonically as $N$ grows.
Using $N{=}500$ yields a relative improvement of 44.24\% in test reduction over $N{=}0$.
Beyond this point, larger values of $N$ bring only marginal additional gains, with $N{=}900$ improving test reduction by only 2.52\% over $N{=}500$.
Different projects saturate at different values of $N$, beyond which further increases provide nearly no returns.
For example, \textit{seaborn} exhibits nearly identical test reduction across all evaluated values of $N$ from 100 to 900.
In contrast, \textit{sklearn} reaches saturation around $N{=}300$, while \textit{sympy} and \textit{sphinx} peak at $N{=}500$ and $N{=}700$, respectively.
Time reduction shows a different trend.
On average, \appname achieves the best time reduction when $N{=}500$.
Although test reduction continues to increase for larger values of $N$, time reduction starts to decrease because the additional runtime overhead of monitoring more functions outweighs the benefit of skipping additional tests.
The best value of $N$ varies across projects.
For instance, \textit{sphinx} achieves a time reduction of 17.57\% at $N{=}700$, which is notably higher than the 9.89\% observed at $N{=}500$.
Across all evaluated values of $N$, including $N{=}0$, \appname consistently outperforms the file-level baselines shown in the figure, indicating that its effectiveness is robust to the choice of $N$. 
In our evaluation, we adopt $N{=}500$ as a uniform setting across all projects to balance effectiveness and runtime overhead while keeping the configuration simple.

\vskip 1mm
\noindent \fbox{
	\parbox{0.95\linewidth}{\textbf{Answer to RQ4}: $N{=}500$ provides a good balance between effectiveness and overhead, yielding near-saturated test reduction and the best average time reduction. Increasing $N$ improves test reduction, but with diminishing returns. While larger $N$ can further reduce tests, the added runtime overhead eventually limits time reduction.}
}

%% file: tables/dataset.tex
\begin{table}[t]
\centering
\setlength{\tabcolsep}{2pt} %
\caption{Dataset summary. Test Files is the average number of test files, Test Time is the full-suite execution time. NNodes and CENodes are the numbers of name nodes and code element nodes, respectively.}
\scalebox{0.8}{
\begin{tabular}{lcccccc}
\toprule
\textbf{Project} & \textbf{Head} & \textbf{kLoC} & \textbf{Test Files} & \textbf{Test Time(s)} & \textbf{NNodes(k)} & \textbf{CENodes(k)} \\
\midrule
sympy          & d854b09 & 426 & 611.4 & 1391.1 & 24.4 & 60.2 \\
sklearn   & cc526ee & 230 & 243.7 & 725.5  & 11.3 & 41.3 \\
matplotlib     & d05b43d & 156 & 102.0   & 431.0  & 10.0 & 39.0 \\
dask           & 8639b6e & 123 & 163.0   & 867.9  & 6.9  & 16.8 \\
xarray         & 97fb90b & 112 & 68.3  & 726.2  & 6.3  & 12.7 \\
sphinx         & eda953e & 106 & 139.2 & 133.0  & 5.2  & 14.5 \\
pylint         & d17bb06 & 74  & 81.0    & 102.7  & 2.9  & 7.1  \\
seaborn        & 2386036 & 35  & 34.1  & 562.2  & 2.6  & 5.6  \\
pvlib   & b4916e1 & 28  & 54.7  & 56.6   & 2.2  & 5.5  \\
loguru         & 940b7cf & 12  & 48.2  & 76.9   & 1.0  & 1.7  \\
\midrule
Avg.           &  & 130 & 154.6 & 507.3 & 7.3 & 20.4 \\
\bottomrule
\end{tabular}}
\label{tab:dataset}
\end{table}

%% file: tables/rq1.tex
\begin{table*}[t]
\centering
\setlength{\tabcolsep}{2.7pt}
\caption{Comparison with baseline RTS techniques (RQ1). GT denotes the test reduction achieved by the ground truth for each project.}
\scalebox{0.77}{
\begin{tabular}{
lcccc|cccc|cccc|c}
\toprule
\multirow{2}{*}{\textbf{Project}}  
& \multicolumn{4}{c|}{\textbf{\appname}}
& \multicolumn{4}{c|}{\textbf{EkstaP}}
& \multicolumn{4}{c|}{\textbf{BabelRTS}}
& \multicolumn{1}{c}{\textbf{GT}} \\
& {\fontsize{9pt}{10pt}\selectfont \textbf{TestR}} 
& {\fontsize{9pt}{10pt}\selectfont \textbf{Precision}} 
& {\fontsize{9pt}{10pt}\selectfont \textbf{TimeR}} 
& {\fontsize{9pt}{10pt}\selectfont \textbf{SafeR}}
& {\fontsize{9pt}{10pt}\selectfont \textbf{TestR}} 
& {\fontsize{9pt}{10pt}\selectfont \textbf{Precision}} 
& {\fontsize{9pt}{10pt}\selectfont \textbf{TimeR}} 
& {\fontsize{9pt}{10pt}\selectfont \textbf{SafeR}}
& {\fontsize{9pt}{10pt}\selectfont \textbf{TestR}} 
& {\fontsize{9pt}{10pt}\selectfont \textbf{Precision}} 
& {\fontsize{9pt}{10pt}\selectfont \textbf{TimeR}} 
& {\fontsize{9pt}{10pt}\selectfont \textbf{SafeR}}
& {\fontsize{9pt}{10pt}\selectfont \textbf{TestR}} \\
\midrule
sympy          & 73.51\% & 45.47\% & 32.66\% & 100.0\% & 0.00\% & 12.05\% & -3.08\% & 100.0\% & 12.86\% & 12.32\% & 14.16\% & 88.0\% & 87.95\% \\
sklearn        & 89.98\% & 84.52\% & 66.81\% & 100.0\% & 15.83\% & 10.06\% & 12.42\% & 100.0\% & 15.74\% & 10.05\% & 13.52\% & 100.0\% & 91.53\% \\
matplotlib            & 54.43\% & 47.98\% & 37.01\% & 100.0\% & 0.00\% & 21.86\% & -3.52\% & 100.0\% & 97.63\% & 46.28\% & 95.71\% & 14.0\%  & 78.14\% \\
dask           & 51.82\% & 63.69\% & 29.15\% & 100.0\% & 0.00\% & 30.69\% & -3.34\% & 100.0\% & 6.70\% & 32.35\%  & 2.51\%  & 98.0\% & 69.31\% \\
xarray         & 66.28\% & 77.06\% & 49.11\% & 100.0\% & 3.93\% & 27.05\% & 0.52\%  & 100.0\% & 5.74\% & 26.80\%  & 6.08\%  & 66.0\% & 74.01\% \\
sphinx         & 50.47\% & 47.35\% & 9.89\% & 98.0\%  & 2.00\% & 23.94\% & -3.15\% & 100.0\% & 10.37\% & 26.02\% & 6.36\%  & 80.0\% & 76.54\%\\
pylint         & 79.75\% & 51.10\% & 30.30\% & 98.0\%  & 6.00\% & 11.06\% & -1.54\% & 100.0\% & 60.84\% & 16.65\% & 21.76\% & 48.0\% & 89.60\% \\
seaborn        & 82.88\% & 98.29\% & 75.68\% & 100.0\% & 26.55\% & 22.91\% & 20.73\% & 100.0\% & 11.96\% & 19.04\% & 5.99\%  & 98.0\% & 83.18\% \\
pvlib          & 94.33\% & 61.29\% & 78.59\% & 100.0\% & 9.90\% & 3.85\% & 5.47\%  & 100.0\% & 46.24\% & 6.39\% & 41.79\% & 98.0\% & 96.53\%\\
loguru         & 55.56\% & 89.72\% & 46.66\% & 100.0\% & 9.97\% & 44.28\% & 6.47\%  & 100.0\% & 15.49\% & 46.58\% & 11.60\% & 76.0\% & 60.13\% \\
\midrule
Avg.           & \textbf{69.90\%} & \textbf{66.65\%} & \textbf{45.59\%} & 99.6\% & 7.42\% & 20.77\% & 3.10\%  & 100.0\% & 28.36\% & 24.25\% & 21.95\% & 76.6\% & 80.69\% \\
\bottomrule
\end{tabular}}
\label{tab:rq1}
\end{table*}

%% file: tables/rq2.tex
\begin{figure*}[t]
\centering

\begin{minipage}[t]{0.50\textwidth}
\vspace{0pt}
\centering
\setlength{\tabcolsep}{4pt}
\resizebox{\linewidth}{!}{
\begin{tabular}{lcccccc}
\toprule
\multirow{2}{*}{\textbf{Project}} &
\multirow{2}{*}{\textbf{RunAll(s)}} &
\multicolumn{5}{c}{\textbf{Time (\% of RunAll)}} \\
\cmidrule{3-7}
& & \textbf{Test} & \textbf{Runtime} & \textbf{Select} & \textbf{Init} & \textbf{Total} \\
\midrule
sympy   & 69555.2 & 46.22\% & 10.64\% & 7.43\% & 3.05\% & 67.34\% \\
sklearn & 36277.0 & 27.04\% & 1.31\%  & 2.33\% & 2.51\% & 33.19\% \\
mpl     & 21548.0 & 53.83\% & 3.92\%  & 2.95\% & 2.30\% & 63.00\% \\
dask    & 43395.8 & 62.73\% & 4.32\%  & 1.58\% & 2.22\% & 70.85\% \\
xarray  & 36311.8 & 41.87\% & 5.72\%  & 1.04\% & 2.25\% & 50.88\% \\
sphinx  & 6647.6  & 66.95\% & 12.15\% & 8.24\% & 2.77\% & 90.11\% \\
pylint  & 5132.8  & 63.12\% & 1.02\%  & 3.35\% & 2.21\% & 69.70\% \\
seaborn & 28110.5 & 21.72\% & 0.20\%  & 0.43\% & 1.96\% & 24.31\% \\
pvlib   & 2832.3  & 12.17\% & 3.56\%  & 3.16\% & 2.52\% & 21.41\% \\
loguru  & 3845.3  & 48.07\% & 1.65\%  & 1.66\% & 1.96\% & 53.34\% \\
\midrule
Avg.    & 25365.6 & 44.37\% & 4.45\% & 3.22\% & 2.38\% & 54.41\% \\
\bottomrule
\end{tabular}}
\captionof{table}{Breakdown of the overhead of \appname relative to full test-suite execution (RQ2). 
}
\label{tab:rq2}
\end{minipage}
\hfill
\begin{minipage}[t]{0.47\textwidth}
\vspace{0pt}
\includegraphics[width=\linewidth]{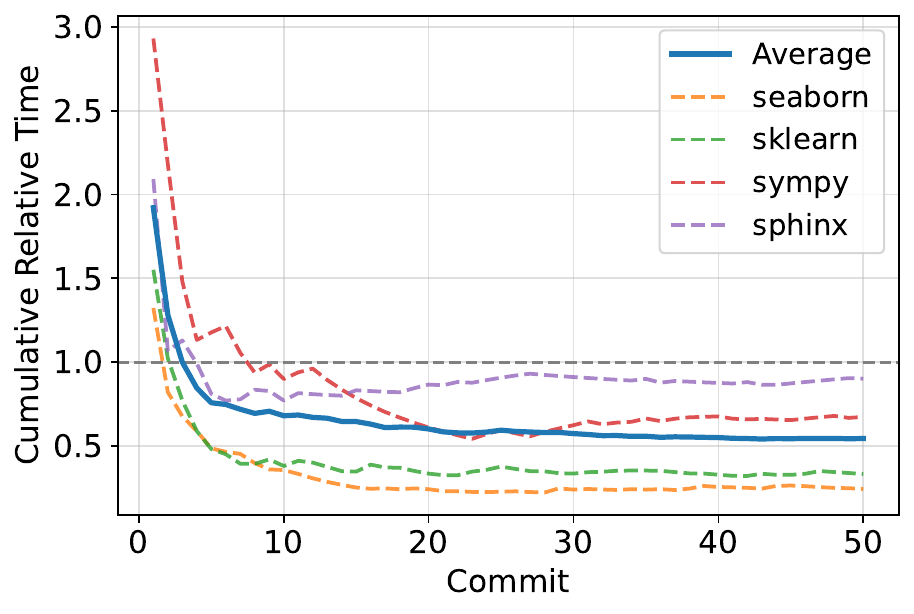}
\captionof{figure}{Cumulative Relative Testing Time}
\label{fig:rq2_cumulative}
\end{minipage}
\end{figure*}

%% file: tables/rq3.tex
\begin{table}[t]
\centering
\setlength{\tabcolsep}{4pt}
\caption{Ablation of pruning mechanisms in \appname (RQ3).}
\scalebox{0.8}{
\begin{tabular}{lcc|cc|cc|cc}
\toprule
\multirow{2}{*}{\textbf{Project}} 
& \multicolumn{2}{c|}{\textbf{\appname}}
& \multicolumn{2}{c|}{\textbf{w/o CF}}
& \multicolumn{2}{c|}{\textbf{w/o NEM}}
& \multicolumn{2}{c}{\textbf{w/o CF\&NEM}} \\
& \textbf{TestR} & \textbf{TimeR}  
& \textbf{TestR} & \textbf{TimeR} 
& \textbf{TestR} & \textbf{TimeR}
& \textbf{TestR} & \textbf{TimeR} \\
\midrule
sympy          & 73.51\% & 32.66\% & 40.99\% & 20.59\% & 73.42\% & 28.15\% & 15.93\% & -19.29\% \\
sklearn        & 89.98\% & 66.81\% & 76.06\% & 57.59\% & 89.92\% & 65.49\% & 75.92\% & 55.63\% \\
mpl            & 54.43\% & 37.01\% & 32.04\% & 18.24\% & 47.55\% & 29.37\% & 30.10\% & 15.98\% \\
dask           & 51.82\% & 29.15\% & 14.15\% & 1.70\%  & 48.80\% & 21.09\% & 8.44\%  & -6.22\% \\
xarray         & 66.28\% & 49.11\% & 37.45\% & 25.99\% & 65.34\% & 46.76\% & 33.55\% & 23.15\% \\
sphinx         & 50.47\% & 9.89\%  & 9.99\%  & -16.53\% & 48.10\% & 6.08\%  & 9.93\%  & -20.84\% \\
pylint         & 79.75\% & 30.30\% & 67.21\% & 16.59\% & 79.41\% & 28.90\% & 66.86\% & 15.10\% \\
seaborn        & 82.88\% & 75.68\% & 81.18\% & 75.53\% & 82.88\% & 75.32\% & 80.01\% & 74.20\% \\
pvlib          & 94.33\% & 78.59\% & 93.86\% & 78.30\% & 92.43\% & 77.85\% & 91.81\% & 78.49\% \\
loguru         & 55.56\% & 46.66\% & 31.69\% & 26.66\% & 55.48\% & 45.89\% & 31.60\% & 26.53\% \\
\midrule
avg.           
& \textbf{69.90\%} & \textbf{45.59\%} & 48.46\% & 30.47\% & 68.33\% & 42.49\% & 44.42\% & 24.27\% \\
\textbf{$\Delta$ vs. \appname} 
&  &  
& \textcolor{red}{30.67\%$\downarrow$} & \textcolor{red}{33.17\%$\downarrow$}
& \textcolor{red}{2.24\%$\downarrow$} & \textcolor{red}{6.79\%$\downarrow$}
& \textcolor{red}{36.46\%$\downarrow$} & \textcolor{red}{46.75\%$\downarrow$} \\
\bottomrule
\end{tabular}}
\label{tab:rq3}
\end{table}

%% file: sections/discussion.tex
\section{Threats to Validity}

\paragraph{Threats to internal validity.}
Our pruning of critical functions relies on runtime invocation information from previous executions to decide whether a function is reachable by a given test.
This use of dynamic information is safe because, before propagation reaches a modified code element, the executed code remains unchanged, and previously observed invocations remain valid. 
Once propagation reaches a modified element, the test is already identified as affected, and any subsequent behavioral changes do not influence the decision.
Our implementation of \appname may not capture all Python language features. 
We address several constructs that influence RTS safety or effectiveness, such as decorators and operator overloading, but less common patterns may not be fully supported.
To reduce this risk, we evaluate \appname on a dataset of 500 commits drawn from ten large projects so that widely used language features are represented. 
\appname is also extensible. 
Additional names can be injected into the \textit{usedNames} sets of individual code elements or into the initial name sets of test files to accommodate project-specific patterns when needed.
\appname does not model fully dynamic code execution mechanisms such as \texttt{eval} and \texttt{exec}.
Such mechanisms are well known to be difficult to analyze efficiently \cite{li2019understanding,aakerblom2014tracing}, and resolving them precisely would counteract the goal of RTS. 
Although this limitation may affect safety in principle, we did not observe any missed tests caused by these mechanisms in our dataset, which suggests that such cases are uncommon in practice.
Future work could explore lightweight ways to approximate reflective behavior without incurring the high cost of full dynamic analysis, enabling safer RTS for reflection-heavy code.

\paragraph{Threats to external validity.}
Our dataset may not fully represent real-world development practices.
To mitigate this threat, we select ten large, popular open-source projects from various application domains.
For each project, we follow prior work~\cite{zhang2024hybrid} and collect fifty consecutive commits. 
While fifty commits may not reflect behavior across the full history of a project, this number balances evaluation cost and generalizability.

%% file: sections/relatedwork.tex
\section{Related Work}

\paragraph{RTS for statically typed languages.}
RTS has been studied most extensively in statically typed languages, particularly Java.
Most existing techniques rely on dependency analysis, which selects tests that depend on modified program components.
Based on the granularity at which dependencies are tracked, these techniques can be broadly categorized into file-level and function-level approaches.
File-level RTS~\cite{gligoric2015practical,legunsen2017starts,vasic2017file} tracks dependencies between test files and source files, and selects a test if any file it depends on has changed.
Ekstazi~\cite{gligoric2015practical} and STARTS~\cite{legunsen2017starts} derive file dependencies through dynamic and static analysis respectively.
Ekstazi\#~\cite{vasic2017file} instruments C\# assemblies to capture file-level dependencies, 
enabling RTS for .NET. 
Our results show that file-level RTS is far less effective for Python because eager importing inflates dependency scopes. 
Function-level RTS~\cite{fu2019resurgence,hundsdorfer2025rustyrts,zhang2018hybrid,wang2018towards,liu2023more,zhang2024hybrid,zaber2021towards,zhu2019framework,soetens2016change,li2019method,shi2019reflection,elsner2023binaryrts} aims to improve precision by tracking dependencies at the level of functions, typically relying on call graph construction to approximate test reachability.
RTS++~\cite{fu2019resurgence} analyzes LLVM IR to build call graphs across revisions and selects tests whose dependency closures reach modified functions.
HyRTS~\cite{zhang2018hybrid} applies function-level analysis selectively to changed files, falling back to file-level analysis elsewhere to balance precision and overhead.
Wang et al.~\cite{wang2018towards} and Liu et al.~\cite{liu2023more} identify semantics-modifying changes and select tests only for such changes, excluding refactorings and other non-semantic edits. 
JcgEks~\cite{zhang2024hybrid} combines dynamic file-level selection with call graph analysis augmented by runtime information to determine whether tests can reach modified methods.
These approaches depend on constructing precise call graphs.
Because Python is dynamically typed, building high-quality call graphs is significantly harder, making such techniques difficult to port directly.
In contrast, \appname avoids call graph construction while still achieving high precision.

\paragraph{RTS for dynamically typed languages and language-agnostic RTS approaches.}
A few studies explore RTS for dynamically typed languages, where precise static dependency analysis is more challenging.
NodeSRT~\cite{chen2021nodesrt} addresses challenges in JavaScript by instrumenting all functions to collect runtime call edges across Node.js and browser environments.
Pytest-rts~\cite{kauhanen2021regression} selects tests based on previously observed coverage at the test-function level, whereas our evaluation operates at the test-file level.
This mismatch prevents a fair comparison, so we do not include pytest-rts as a baseline.
Compared to NodeSRT and pytest-rts, \appname instruments only critical functions, striking a balance between test reduction and runtime overhead.
BabelRTS~\cite{maurina2025babelrts} performs language-agnostic RTS via regex-based rules to extract file-level dependencies.
Other work explores language-agnostic learning-based RTS~\cite{anderson2014improving,azizi2018retest,machalica2019predictive,al2020selective,bertolino2020learning} that predicts affected tests from historical features, such as coverage or failure patterns, or predicts when to entirely skip regression testing~\cite{issta2021-tests}.
These methods do not offer safety guarantees, whereas \appname focuses on analysis-based selection with minimal safety loss.

%% file: sections/conclusion.tex
\section{Conclusion}

This paper presents \appname, a regression test selection approach for Python that addresses the key challenges of dependency analysis in Python RTS. 
\appname is built on \emph{name-based dependency propagation}, a conservative dependency analysis framework based on direct name-element matching, which deliberately over-approximates dependencies to ensure safety. 
This design reflects the fundamental requirement of RTS to remain safe, especially in dynamic languages where precise static dependency analysis is inherently difficult to achieve.
On top of this foundation, \appname applies two pruning strategies to eliminate false dependencies introduced by coarse name-element matching, improving precision while preserving safety. 
To enable rigorous evaluation, we curate the first Python RTS dataset with a precise ground truth.
Experimental results show that \appname is substantially more effective than existing Python RTS approaches, advancing the practicality of regression test selection for Python.

%% file: main.bbl

\begin{thebibliography}{58}

%%% the Web address.

\ifx \showCODEN    \undefined \def \showCODEN     #1{\unskip}     \fi
\ifx \showISBNx    \undefined \def \showISBNx     #1{\unskip}     \fi
\ifx \showISBNxiii \undefined \def \showISBNxiii  #1{\unskip}     \fi
\ifx \showISSN     \undefined \def \showISSN      #1{\unskip}     \fi
\ifx \showLCCN     \undefined \def \showLCCN      #1{\unskip}     \fi
\ifx \shownote     \undefined \def \shownote      #1{#1}          \fi
\ifx \showarticletitle \undefined \def \showarticletitle #1{#1}   \fi
\ifx \showURL      \undefined \def \showURL       {\relax}        \fi
% invisible to TeX
\providecommand\bibfield[2]{#2}
\providecommand\bibinfo[2]{#2}
\providecommand\natexlab[1]{#1}
\providecommand\showeprint[2][]{arXiv:#2}

\bibitem[pyt(2025a)]%
        {pythondatamodel}
 \bibinfo{year}{2025}\natexlab{a}.
\newblock \bibinfo{title}{3. Data model - Python 3.14.0 documentation}.
\newblock
\urldef\tempurl%
\url{https://docs.python.org/3/reference/datamodel.html}
\showURL{%
\tempurl}


\bibitem[pyt(2025b)]%
        {pythondecorator}
 \bibinfo{year}{2025}\natexlab{b}.
\newblock \bibinfo{title}{Compound statements -- Python 3.14.0 documentation}.
\newblock
\urldef\tempurl%
\url{https://docs.python.org/3/reference/compound_stmts.html#function-definitions}
\showURL{%
\tempurl}


\bibitem[pyt(2025c)]%
        {pythonhasname}
 \bibinfo{year}{2025}\natexlab{c}.
\newblock \bibinfo{title}{dis - Disassembler for Python bytecode - Python 3.14.0 documentation}.
\newblock
\urldef\tempurl%
\url{https://docs.python.org/3/library/dis.html#dis.hasname}
\showURL{%
\tempurl}


\bibitem[pyt(2025d)]%
        {pythonimport}
 \bibinfo{year}{2025}\natexlab{d}.
\newblock \bibinfo{title}{The import system -- Python 3.14.0 documentation}.
\newblock
\urldef\tempurl%
\url{https://docs.python.org/3/reference/import.html#regular-packages}
\showURL{%
\tempurl}


\bibitem[git(2025)]%
        {githubpython}
 \bibinfo{year}{2025}\natexlab{}.
\newblock \bibinfo{title}{Octoverse: A new developer joins GitHub every second as AI leads TypeScript to \#1}.
\newblock
\urldef\tempurl%
\url{https://github.blog/news-insights/octoverse/octoverse-a-new-developer-joins-github-every-second-as-ai-leads-typescript-to-1/}
\showURL{%
\tempurl}


\bibitem[sta(2025)]%
        {stackoverflowpython}
 \bibinfo{year}{2025}\natexlab{}.
\newblock \bibinfo{title}{Technology | 2025 Stack Overflow Developer Survey}.
\newblock
\urldef\tempurl%
\url{https://survey.stackoverflow.co/2025/technology#most-popular-technologies}
\showURL{%
\tempurl}


\bibitem[tio(2025)]%
        {tiobepython}
 \bibinfo{year}{2025}\natexlab{}.
\newblock \bibinfo{title}{TIOBE Index - TIOBE}.
\newblock
\urldef\tempurl%
\url{https://www.tiobe.com/tiobe-index/}
\showURL{%
\tempurl}


\bibitem[jet(2025)]%
        {jetbrainpython}
 \bibinfo{year}{2025}\natexlab{}.
\newblock \bibinfo{title}{Tools and Trends - The State of Developer Ecosystem in 2025}.
\newblock
\urldef\tempurl%
\url{https://devecosystem-2025.jetbrains.com/tools-and-trends}
\showURL{%
\tempurl}


\bibitem[pac(2026)]%
        {package}
 \bibinfo{year}{2026}\natexlab{}.
\newblock \bibinfo{title}{Our replication package}.
\newblock
\urldef\tempurl%
\url{https://github.com/ZJU-CTAG/NameRTS}
\showURL{%
\tempurl}


\bibitem[{\AA}kerblom et~al\mbox{.}(2014)]%
        {aakerblom2014tracing}
\bibfield{author}{\bibinfo{person}{Beatrice {\AA}kerblom}, \bibinfo{person}{Jonathan Stendahl}, \bibinfo{person}{Mattias Tumlin}, {and} \bibinfo{person}{Tobias Wrigstad}.} \bibinfo{year}{2014}\natexlab{}.
\newblock \showarticletitle{Tracing dynamic features in python programs}. In \bibinfo{booktitle}{\emph{Proceedings of the 11th working conference on mining software repositories}}. \bibinfo{pages}{292--295}.
\newblock


\bibitem[Al-Sabbagh et~al\mbox{.}(2020)]%
        {al2020selective}
\bibfield{author}{\bibinfo{person}{Khaled~Walid Al-Sabbagh}, \bibinfo{person}{Miroslaw Staron}, \bibinfo{person}{Miroslaw Ochodek}, \bibinfo{person}{Regina Hebig}, {and} \bibinfo{person}{Wilhelm Meding}.} \bibinfo{year}{2020}\natexlab{}.
\newblock \showarticletitle{Selective regression testing based on big data: Comparing feature extraction techniques}. In \bibinfo{booktitle}{\emph{2020 IEEE International Conference on Software Testing, Verification and Validation Workshops}}. \bibinfo{pages}{322--329}.
\newblock


\bibitem[Anderson et~al\mbox{.}(2014)]%
        {anderson2014improving}
\bibfield{author}{\bibinfo{person}{Jeff Anderson}, \bibinfo{person}{Saeed Salem}, {and} \bibinfo{person}{Hyunsook Do}.} \bibinfo{year}{2014}\natexlab{}.
\newblock \showarticletitle{Improving the effectiveness of test suite through mining historical data}. In \bibinfo{booktitle}{\emph{Proceedings of the 11th Working Conference on Mining Software Repositories}}. \bibinfo{pages}{142--151}.
\newblock


\bibitem[Azizi and Do(2018)]%
        {azizi2018retest}
\bibfield{author}{\bibinfo{person}{Maral Azizi} {and} \bibinfo{person}{Hyunsook Do}.} \bibinfo{year}{2018}\natexlab{}.
\newblock \showarticletitle{ReTEST: A cost effective test case selection technique for modern software development}. In \bibinfo{booktitle}{\emph{2018 IEEE 29th International Symposium on Software Reliability Engineering}}. \bibinfo{pages}{144--154}.
\newblock


\bibitem[Bertolino et~al\mbox{.}(2020)]%
        {bertolino2020learning}
\bibfield{author}{\bibinfo{person}{Antonia Bertolino}, \bibinfo{person}{Antonio Guerriero}, \bibinfo{person}{Breno Miranda}, \bibinfo{person}{Roberto Pietrantuono}, {and} \bibinfo{person}{Stefano Russo}.} \bibinfo{year}{2020}\natexlab{}.
\newblock \showarticletitle{Learning-to-rank vs ranking-to-learn: Strategies for regression testing in continuous integration}. In \bibinfo{booktitle}{\emph{Proceedings of the ACM/IEEE 42nd International Conference on Software Engineering}}. \bibinfo{pages}{1--12}.
\newblock


\bibitem[Blondeau et~al\mbox{.}(2017)]%
        {blondeau2017test}
\bibfield{author}{\bibinfo{person}{Vincent Blondeau}, \bibinfo{person}{Anne Etien}, \bibinfo{person}{Nicolas Anquetil}, \bibinfo{person}{Sylvain Cresson}, \bibinfo{person}{Pascal Croisy}, {and} \bibinfo{person}{St{\'e}phane Ducasse}.} \bibinfo{year}{2017}\natexlab{}.
\newblock \showarticletitle{Test case selection in industry: An analysis of issues related to static approaches}.
\newblock \bibinfo{journal}{\emph{Software Quality Journal}} \bibinfo{volume}{25}, \bibinfo{number}{4} (\bibinfo{year}{2017}), \bibinfo{pages}{1203--1237}.
\newblock


\bibitem[Bouzenia et~al\mbox{.}(2024)]%
        {bouzenia2024dypybench}
\bibfield{author}{\bibinfo{person}{Islem Bouzenia}, \bibinfo{person}{Bajaj~Piyush Krishan}, {and} \bibinfo{person}{Michael Pradel}.} \bibinfo{year}{2024}\natexlab{}.
\newblock \showarticletitle{DyPyBench: A benchmark of executable python software}.
\newblock \bibinfo{journal}{\emph{Proceedings of the ACM on Software Engineering}}  \bibinfo{volume}{1} (\bibinfo{year}{2024}), \bibinfo{pages}{338--358}.
\newblock


\bibitem[Bouzenia and Pradel(2024)]%
        {bouzenia2024resource}
\bibfield{author}{\bibinfo{person}{Islem Bouzenia} {and} \bibinfo{person}{Michael Pradel}.} \bibinfo{year}{2024}\natexlab{}.
\newblock \showarticletitle{Resource usage and optimization opportunities in workflows of github actions}. In \bibinfo{booktitle}{\emph{Proceedings of the 46th IEEE/ACM International Conference on Software Engineering}}. \bibinfo{pages}{1--12}.
\newblock


\bibitem[Chen(2021)]%
        {chen2021nodesrt}
\bibfield{author}{\bibinfo{person}{Yufeng Chen}.} \bibinfo{year}{2021}\natexlab{}.
\newblock \showarticletitle{NodeSRT: a selective regression testing tool for Node. js application}. In \bibinfo{booktitle}{\emph{2021 IEEE/ACM 43rd International Conference on Software Engineering: Companion Proceedings}}. \bibinfo{pages}{126--128}.
\newblock


\bibitem[Chittimalli and Harrold(2009)]%
        {chittimalli2009recomputing}
\bibfield{author}{\bibinfo{person}{Pavan~Kumar Chittimalli} {and} \bibinfo{person}{Mary~Jean Harrold}.} \bibinfo{year}{2009}\natexlab{}.
\newblock \showarticletitle{Recomputing coverage information to assist regression testing}.
\newblock \bibinfo{journal}{\emph{IEEE Transactions on Software Engineering}} \bibinfo{volume}{35}, \bibinfo{number}{4} (\bibinfo{year}{2009}), \bibinfo{pages}{452--469}.
\newblock


\bibitem[Deng et~al\mbox{.}(2025)]%
        {deng2025nocode}
\bibfield{author}{\bibinfo{person}{Le Deng}, \bibinfo{person}{Zhonghao Jiang}, \bibinfo{person}{Jialun Cao}, \bibinfo{person}{Michael Pradel}, {and} \bibinfo{person}{Zhongxin Liu}.} \bibinfo{year}{2025}\natexlab{}.
\newblock \showarticletitle{NoCode-bench: A Benchmark for Evaluating Natural Language-Driven Feature Addition}.
\newblock \bibinfo{journal}{\emph{CoRR}}  \bibinfo{volume}{abs/2507.18130} (\bibinfo{year}{2025}).
\newblock


\bibitem[Eghbali and Pradel(2022)]%
        {eghbali2022dynapyt}
\bibfield{author}{\bibinfo{person}{Aryaz Eghbali} {and} \bibinfo{person}{Michael Pradel}.} \bibinfo{year}{2022}\natexlab{}.
\newblock \showarticletitle{DynaPyt: a dynamic analysis framework for Python}. In \bibinfo{booktitle}{\emph{Proceedings of the 30th ACM Joint European Software Engineering Conference and Symposium on the Foundations of Software Engineering}}. \bibinfo{pages}{760--771}.
\newblock


\bibitem[Elsner et~al\mbox{.}(2023)]%
        {elsner2023binaryrts}
\bibfield{author}{\bibinfo{person}{Daniel Elsner}, \bibinfo{person}{Severin Kacianka}, \bibinfo{person}{Stephan Lipp}, \bibinfo{person}{Alexander Pretschner}, \bibinfo{person}{Axel Habermann}, \bibinfo{person}{Maria Graber}, {and} \bibinfo{person}{Silke Reimer}.} \bibinfo{year}{2023}\natexlab{}.
\newblock \showarticletitle{BinaryRTS: Cross-language regression test selection for C++ binaries in CI}. In \bibinfo{booktitle}{\emph{2023 IEEE Conference on Software Testing, Verification and Validation}}. \bibinfo{pages}{327--338}.
\newblock


\bibitem[Engstr{\"o}m et~al\mbox{.}(2010)]%
        {engstrom2010systematic}
\bibfield{author}{\bibinfo{person}{Emelie Engstr{\"o}m}, \bibinfo{person}{Per Runeson}, {and} \bibinfo{person}{Mats Skoglund}.} \bibinfo{year}{2010}\natexlab{}.
\newblock \showarticletitle{A systematic review on regression test selection techniques}.
\newblock \bibinfo{journal}{\emph{Information and Software Technology}} \bibinfo{volume}{52}, \bibinfo{number}{1} (\bibinfo{year}{2010}), \bibinfo{pages}{14--30}.
\newblock


\bibitem[Fu et~al\mbox{.}(2019)]%
        {fu2019resurgence}
\bibfield{author}{\bibinfo{person}{Ben Fu}, \bibinfo{person}{Sasa Misailovic}, {and} \bibinfo{person}{Milos Gligoric}.} \bibinfo{year}{2019}\natexlab{}.
\newblock \showarticletitle{Resurgence of regression test selection for C++}. In \bibinfo{booktitle}{\emph{2019 12th IEEE Conference on Software Testing, Validation and Verification}}. \bibinfo{pages}{323--334}.
\newblock


\bibitem[Gligoric et~al\mbox{.}(2015)]%
        {gligoric2015practical}
\bibfield{author}{\bibinfo{person}{Milos Gligoric}, \bibinfo{person}{Lamyaa Eloussi}, {and} \bibinfo{person}{Darko Marinov}.} \bibinfo{year}{2015}\natexlab{}.
\newblock \showarticletitle{Practical regression test selection with dynamic file dependencies}. In \bibinfo{booktitle}{\emph{Proceedings of the 2015 International Symposium on Software Testing and Analysis}}. \bibinfo{pages}{211--222}.
\newblock


\bibitem[Gyori et~al\mbox{.}(2018)]%
        {gyori2018evaluating}
\bibfield{author}{\bibinfo{person}{Alex Gyori}, \bibinfo{person}{Owolabi Legunsen}, \bibinfo{person}{Farah Hariri}, {and} \bibinfo{person}{Darko Marinov}.} \bibinfo{year}{2018}\natexlab{}.
\newblock \showarticletitle{Evaluating regression test selection opportunities in a very large open-source ecosystem}. In \bibinfo{booktitle}{\emph{2018 IEEE 29th International Symposium on Software Reliability Engineering}}. \bibinfo{pages}{112--122}.
\newblock


\bibitem[Harrold et~al\mbox{.}(1993)]%
        {harrold1993methodology}
\bibfield{author}{\bibinfo{person}{M~Jean Harrold}, \bibinfo{person}{Rajiv Gupta}, {and} \bibinfo{person}{Mary~Lou Soffa}.} \bibinfo{year}{1993}\natexlab{}.
\newblock \showarticletitle{A methodology for controlling the size of a test suite}.
\newblock \bibinfo{journal}{\emph{ACM Transactions on Software Engineering and Methodology}} \bibinfo{volume}{2}, \bibinfo{number}{3} (\bibinfo{year}{1993}), \bibinfo{pages}{270--285}.
\newblock


\bibitem[Hundsdorfer et~al\mbox{.}(2025)]%
        {hundsdorfer2025rustyrts}
\bibfield{author}{\bibinfo{person}{Simon Hundsdorfer}, \bibinfo{person}{Roland W{\"u}rsching}, {and} \bibinfo{person}{Alexander Pretschner}.} \bibinfo{year}{2025}\natexlab{}.
\newblock \showarticletitle{RustyRTS: Regression Test Selection for Rust}. In \bibinfo{booktitle}{\emph{2025 IEEE Conference on Software Testing, Verification and Validation}}. \bibinfo{pages}{338--348}.
\newblock


\bibitem[Jiang et~al\mbox{.}(2025)]%
        {jiang2025agentic}
\bibfield{author}{\bibinfo{person}{Zhonghao Jiang}, \bibinfo{person}{David Lo}, {and} \bibinfo{person}{Zhongxin Liu}.} \bibinfo{year}{2025}\natexlab{}.
\newblock \showarticletitle{Agentic Software Issue Resolution with Large Language Models: A Survey}.
\newblock \bibinfo{journal}{\emph{CoRR}}  \bibinfo{volume}{abs/2512.22256} (\bibinfo{year}{2025}).
\newblock


\bibitem[Jimenez et~al\mbox{.}(2024)]%
        {jimenez2023swe}
\bibfield{author}{\bibinfo{person}{Carlos~E Jimenez}, \bibinfo{person}{John Yang}, \bibinfo{person}{Alexander Wettig}, \bibinfo{person}{Shunyu Yao}, \bibinfo{person}{Kexin Pei}, \bibinfo{person}{Ofir Press}, {and} \bibinfo{person}{Karthik Narasimhan}.} \bibinfo{year}{2024}\natexlab{}.
\newblock \showarticletitle{SWE-bench: Can language models resolve real-world GitHub issues?}. In \bibinfo{booktitle}{\emph{International Conference on Learning Representations}}.
\newblock


\bibitem[Kauhanen et~al\mbox{.}(2021)]%
        {kauhanen2021regression}
\bibfield{author}{\bibinfo{person}{Eero Kauhanen}, \bibinfo{person}{Jukka~K Nurminen}, \bibinfo{person}{Tommi Mikkonen}, {and} \bibinfo{person}{Matvei Pashkovskiy}.} \bibinfo{year}{2021}\natexlab{}.
\newblock \showarticletitle{Regression test selection tool for python in continuous integration process}. In \bibinfo{booktitle}{\emph{2021 IEEE International Conference on Software Analysis, Evolution and Reengineering}}. \bibinfo{pages}{618--621}.
\newblock


\bibitem[Law and Rothermel(2003)]%
        {law2003whole}
\bibfield{author}{\bibinfo{person}{James Law} {and} \bibinfo{person}{Gregg Rothermel}.} \bibinfo{year}{2003}\natexlab{}.
\newblock \showarticletitle{Whole program path-based dynamic impact analysis}. In \bibinfo{booktitle}{\emph{25th International Conference on Software Engineering, 2003. Proceedings.}} \bibinfo{pages}{308--318}.
\newblock


\bibitem[Legunsen et~al\mbox{.}(2016)]%
        {legunsen2016extensive}
\bibfield{author}{\bibinfo{person}{Owolabi Legunsen}, \bibinfo{person}{Farah Hariri}, \bibinfo{person}{August Shi}, \bibinfo{person}{Yafeng Lu}, \bibinfo{person}{Lingming Zhang}, {and} \bibinfo{person}{Darko Marinov}.} \bibinfo{year}{2016}\natexlab{}.
\newblock \showarticletitle{An extensive study of static regression test selection in modern software evolution}. In \bibinfo{booktitle}{\emph{Proceedings of the 2016 24th ACM SIGSOFT International Symposium on Foundations of Software Engineering}}. \bibinfo{pages}{583--594}.
\newblock


\bibitem[Legunsen et~al\mbox{.}(2017)]%
        {legunsen2017starts}
\bibfield{author}{\bibinfo{person}{Owolabi Legunsen}, \bibinfo{person}{August Shi}, {and} \bibinfo{person}{Darko Marinov}.} \bibinfo{year}{2017}\natexlab{}.
\newblock \showarticletitle{STARTS: STAtic regression test selection}. In \bibinfo{booktitle}{\emph{2017 32nd IEEE/ACM International Conference on Automated Software Engineering}}. \bibinfo{pages}{949--954}.
\newblock


\bibitem[Leung and White(1989)]%
        {leung1989insights}
\bibfield{author}{\bibinfo{person}{Hareton~KN Leung} {and} \bibinfo{person}{Lee White}.} \bibinfo{year}{1989}\natexlab{}.
\newblock \showarticletitle{Insights into regression testing (software testing)}. In \bibinfo{booktitle}{\emph{Proceedings. Conference on Software Maintenance-1989}}. \bibinfo{pages}{60--69}.
\newblock


\bibitem[Leung and White(1990)]%
        {leung1990study}
\bibfield{author}{\bibinfo{person}{Hareton~KN Leung} {and} \bibinfo{person}{Lee White}.} \bibinfo{year}{1990}\natexlab{}.
\newblock \showarticletitle{A study of integration testing and software regression at the integration level}. In \bibinfo{booktitle}{\emph{Proceedings. Conference on Software Maintenance 1990}}. \bibinfo{pages}{290--301}.
\newblock


\bibitem[Li et~al\mbox{.}(2019a)]%
        {li2019understanding}
\bibfield{author}{\bibinfo{person}{Yue Li}, \bibinfo{person}{Tian Tan}, {and} \bibinfo{person}{Jingling Xue}.} \bibinfo{year}{2019}\natexlab{a}.
\newblock \showarticletitle{Understanding and analyzing java reflection}.
\newblock \bibinfo{journal}{\emph{ACM Transactions on Software Engineering and Methodology}} \bibinfo{volume}{28}, \bibinfo{number}{2} (\bibinfo{year}{2019}), \bibinfo{pages}{1--50}.
\newblock


\bibitem[Li et~al\mbox{.}(2019b)]%
        {li2019method}
\bibfield{author}{\bibinfo{person}{Yingling Li}, \bibinfo{person}{Junjie Wang}, \bibinfo{person}{Yun Yang}, {and} \bibinfo{person}{Qing Wang}.} \bibinfo{year}{2019}\natexlab{b}.
\newblock \showarticletitle{Method-level test selection for continuous integration with static dependencies and dynamic execution rules}. In \bibinfo{booktitle}{\emph{2019 IEEE 19th International Conference on Software Quality, Reliability and Security}}. \bibinfo{pages}{350--361}.
\newblock


\bibitem[Liu et~al\mbox{.}(2023)]%
        {liu2023more}
\bibfield{author}{\bibinfo{person}{Yu Liu}, \bibinfo{person}{Jiyang Zhang}, \bibinfo{person}{Pengyu Nie}, \bibinfo{person}{Milos Gligoric}, {and} \bibinfo{person}{Owolabi Legunsen}.} \bibinfo{year}{2023}\natexlab{}.
\newblock \showarticletitle{More precise regression test selection via reasoning about semantics-modifying changes}. In \bibinfo{booktitle}{\emph{Proceedings of the 32nd ACM SIGSOFT International Symposium on Software Testing and Analysis}}. \bibinfo{pages}{664--676}.
\newblock


\bibitem[Machalica et~al\mbox{.}(2019)]%
        {machalica2019predictive}
\bibfield{author}{\bibinfo{person}{Mateusz Machalica}, \bibinfo{person}{Alex Samylkin}, \bibinfo{person}{Meredith Porth}, {and} \bibinfo{person}{Satish Chandra}.} \bibinfo{year}{2019}\natexlab{}.
\newblock \showarticletitle{Predictive test selection}. In \bibinfo{booktitle}{\emph{2019 IEEE/ACM 41st International Conference on Software Engineering: Software Engineering in Practice}}. \bibinfo{pages}{91--100}.
\newblock


\bibitem[Maurina et~al\mbox{.}(2025)]%
        {maurina2025babelrts}
\bibfield{author}{\bibinfo{person}{Gabriele Maurina}, \bibinfo{person}{Walter Cazzola}, {and} \bibinfo{person}{Sudipto Ghosh}.} \bibinfo{year}{2025}\natexlab{}.
\newblock \showarticletitle{BabelRTS: Polyglot Regression Test Selection}.
\newblock \bibinfo{journal}{\emph{IEEE Transactions on Software Engineering}} (\bibinfo{year}{2025}).
\newblock


\bibitem[Orso et~al\mbox{.}(2004)]%
        {orso2004scaling}
\bibfield{author}{\bibinfo{person}{Alessandro Orso}, \bibinfo{person}{Nanjuan Shi}, {and} \bibinfo{person}{Mary~Jean Harrold}.} \bibinfo{year}{2004}\natexlab{}.
\newblock \showarticletitle{Scaling regression testing to large software systems}.
\newblock \bibinfo{journal}{\emph{ACM SIGSOFT Software Engineering Notes}} \bibinfo{volume}{29}, \bibinfo{number}{6} (\bibinfo{year}{2004}), \bibinfo{pages}{241--251}.
\newblock


\bibitem[Pan and Pradel(2021)]%
        {issta2021-tests}
\bibfield{author}{\bibinfo{person}{Cong Pan} {and} \bibinfo{person}{Michael Pradel}.} \bibinfo{year}{2021}\natexlab{}.
\newblock \showarticletitle{Continuous test suite failure prediction}. In \bibinfo{booktitle}{\emph{Proceedings of the 30th ACM SIGSOFT International Symposium on Software Testing and Analysis}}. \bibinfo{pages}{553--565}.
\newblock


\bibitem[Rothermel and Harrold(1997)]%
        {rothermel1997safe}
\bibfield{author}{\bibinfo{person}{Gregg Rothermel} {and} \bibinfo{person}{Mary~Jean Harrold}.} \bibinfo{year}{1997}\natexlab{}.
\newblock \showarticletitle{A safe, efficient regression test selection technique}.
\newblock \bibinfo{journal}{\emph{ACM Transactions on Software Engineering and Methodology}} \bibinfo{volume}{6}, \bibinfo{number}{2} (\bibinfo{year}{1997}), \bibinfo{pages}{173--210}.
\newblock


\bibitem[Rothermel and Harrold(2002)]%
        {rothermel2002analyzing}
\bibfield{author}{\bibinfo{person}{Gregg Rothermel} {and} \bibinfo{person}{Mary~Jean Harrold}.} \bibinfo{year}{2002}\natexlab{}.
\newblock \showarticletitle{Analyzing regression test selection techniques}.
\newblock \bibinfo{journal}{\emph{IEEE Transactions on software engineering}} \bibinfo{volume}{22}, \bibinfo{number}{8} (\bibinfo{year}{2002}), \bibinfo{pages}{529--551}.
\newblock


\bibitem[Salis et~al\mbox{.}(2021)]%
        {salis2021pycg}
\bibfield{author}{\bibinfo{person}{Vitalis Salis}, \bibinfo{person}{Thodoris Sotiropoulos}, \bibinfo{person}{Panos Louridas}, \bibinfo{person}{Diomidis Spinellis}, {and} \bibinfo{person}{Dimitris Mitropoulos}.} \bibinfo{year}{2021}\natexlab{}.
\newblock \showarticletitle{Pycg: Practical call graph generation in python}. In \bibinfo{booktitle}{\emph{2021 IEEE/ACM 43rd International Conference on Software Engineering}}. \bibinfo{pages}{1646--1657}.
\newblock


\bibitem[Shi et~al\mbox{.}(2019)]%
        {shi2019reflection}
\bibfield{author}{\bibinfo{person}{August Shi}, \bibinfo{person}{Milica Hadzi-Tanovic}, \bibinfo{person}{Lingming Zhang}, \bibinfo{person}{Darko Marinov}, {and} \bibinfo{person}{Owolabi Legunsen}.} \bibinfo{year}{2019}\natexlab{}.
\newblock \showarticletitle{Reflection-aware static regression test selection}.
\newblock \bibinfo{journal}{\emph{Proceedings of the ACM on Programming Languages}}  \bibinfo{volume}{3} (\bibinfo{year}{2019}), \bibinfo{pages}{1--29}.
\newblock


\bibitem[Soetens et~al\mbox{.}(2016)]%
        {soetens2016change}
\bibfield{author}{\bibinfo{person}{Quinten~David Soetens}, \bibinfo{person}{Serge Demeyer}, \bibinfo{person}{Andy Zaidman}, {and} \bibinfo{person}{Javier P{\'e}rez}.} \bibinfo{year}{2016}\natexlab{}.
\newblock \showarticletitle{Change-based test selection: an empirical evaluation}.
\newblock \bibinfo{journal}{\emph{Empirical software engineering}} \bibinfo{volume}{21}, \bibinfo{number}{5} (\bibinfo{year}{2016}), \bibinfo{pages}{1990--2032}.
\newblock


\bibitem[Vasic et~al\mbox{.}(2017)]%
        {vasic2017file}
\bibfield{author}{\bibinfo{person}{Marko Vasic}, \bibinfo{person}{Zuhair Parvez}, \bibinfo{person}{Aleksandar Milicevic}, {and} \bibinfo{person}{Milos Gligoric}.} \bibinfo{year}{2017}\natexlab{}.
\newblock \showarticletitle{File-level vs. module-level regression test selection for. net}. In \bibinfo{booktitle}{\emph{Proceedings of the 2017 11th Joint Meeting on Foundations of Software Engineering}}. \bibinfo{pages}{848--853}.
\newblock


\bibitem[Wang et~al\mbox{.}(2018)]%
        {wang2018towards}
\bibfield{author}{\bibinfo{person}{Kaiyuan Wang}, \bibinfo{person}{Chenguang Zhu}, \bibinfo{person}{Ahmet Celik}, \bibinfo{person}{Jongwook Kim}, \bibinfo{person}{Don Batory}, {and} \bibinfo{person}{Milos Gligoric}.} \bibinfo{year}{2018}\natexlab{}.
\newblock \showarticletitle{Towards refactoring-aware regression test selection}. In \bibinfo{booktitle}{\emph{Proceedings of the 40th international conference on software engineering}}. \bibinfo{pages}{233--244}.
\newblock


\bibitem[Wang et~al\mbox{.}(2026)]%
        {wang2025solved}
\bibfield{author}{\bibinfo{person}{You Wang}, \bibinfo{person}{Michael Pradel}, {and} \bibinfo{person}{Zhongxin Liu}.} \bibinfo{year}{2026}\natexlab{}.
\newblock \showarticletitle{Are ``Solved Issues'' in SWE-bench Really Solved Correctly? An Empirical Study}. In \bibinfo{booktitle}{\emph{2026 IEEE/ACM 48th International Conference on Software Engineering}}.
\newblock


\bibitem[Wong et~al\mbox{.}(1997)]%
        {wong1997study}
\bibfield{author}{\bibinfo{person}{W~Eric Wong}, \bibinfo{person}{Joseph~R Horgan}, \bibinfo{person}{Saul London}, {and} \bibinfo{person}{Hiralal Agrawal}.} \bibinfo{year}{1997}\natexlab{}.
\newblock \showarticletitle{A study of effective regression testing in practice}. In \bibinfo{booktitle}{\emph{PROCEEDINGS The Eighth International Symposium On Software Reliability Engineering}}. \bibinfo{pages}{264--274}.
\newblock


\bibitem[Yoo and Harman(2012)]%
        {yoo2012regression}
\bibfield{author}{\bibinfo{person}{Shin Yoo} {and} \bibinfo{person}{Mark Harman}.} \bibinfo{year}{2012}\natexlab{}.
\newblock \showarticletitle{Regression testing minimization, selection and prioritization: a survey}.
\newblock \bibinfo{journal}{\emph{Software testing, verification and reliability}} \bibinfo{volume}{22}, \bibinfo{number}{2} (\bibinfo{year}{2012}), \bibinfo{pages}{67--120}.
\newblock


\bibitem[Zaber(2021)]%
        {zaber2021towards}
\bibfield{author}{\bibinfo{person}{Maruf~Hasan Zaber}.} \bibinfo{year}{2021}\natexlab{}.
\newblock \emph{\bibinfo{title}{Towards Parallelization of Regression Test Selection}}.
\newblock \bibinfo{thesistype}{Master's\ thesis}. \bibinfo{school}{University of California, Irvine}.
\newblock


\bibitem[Zhang et~al\mbox{.}(2026)]%
        {zhang2026can}
\bibfield{author}{\bibinfo{person}{Chengming Zhang}, \bibinfo{person}{Haoye Wang}, \bibinfo{person}{Chuyang Xu}, \bibinfo{person}{Jiakun Liu}, \bibinfo{person}{Kui Liu}, {and} \bibinfo{person}{Zhongxin Liu}.} \bibinfo{year}{2026}\natexlab{}.
\newblock \showarticletitle{Can test cases generated by large language models facilitate automated program repair?}
\newblock \bibinfo{journal}{\emph{Empirical Software Engineering}} \bibinfo{volume}{31}, \bibinfo{number}{3} (\bibinfo{year}{2026}), \bibinfo{pages}{68}.
\newblock


\bibitem[Zhang et~al\mbox{.}(2024)]%
        {zhang2024hybrid}
\bibfield{author}{\bibinfo{person}{Guofeng Zhang}, \bibinfo{person}{Luyao Liu}, \bibinfo{person}{Zhenbang Chen}, {and} \bibinfo{person}{Ji Wang}.} \bibinfo{year}{2024}\natexlab{}.
\newblock \showarticletitle{Hybrid Regression Test Selection by Integrating File and Method Dependences}. In \bibinfo{booktitle}{\emph{Proceedings of the 39th IEEE/ACM International Conference on Automated Software Engineering}}. \bibinfo{pages}{1557--1569}.
\newblock


\bibitem[Zhang(2018)]%
        {zhang2018hybrid}
\bibfield{author}{\bibinfo{person}{Lingming Zhang}.} \bibinfo{year}{2018}\natexlab{}.
\newblock \showarticletitle{Hybrid regression test selection}. In \bibinfo{booktitle}{\emph{Proceedings of the 40th International Conference on Software Engineering}}. \bibinfo{pages}{199--209}.
\newblock


\bibitem[Zhu et~al\mbox{.}(2019)]%
        {zhu2019framework}
\bibfield{author}{\bibinfo{person}{Chenguang Zhu}, \bibinfo{person}{Owolabi Legunsen}, \bibinfo{person}{August Shi}, {and} \bibinfo{person}{Milos Gligoric}.} \bibinfo{year}{2019}\natexlab{}.
\newblock \showarticletitle{A framework for checking regression test selection tools}. In \bibinfo{booktitle}{\emph{2019 IEEE/ACM 41st International Conference on Software Engineering}}. \bibinfo{pages}{430--441}.
\newblock


\end{thebibliography}
